\documentclass[twocolumn,aps,showpacs,amsmath,amssymb,prb]{revtex4}
\usepackage{epsfig}
\usepackage{graphics}
\usepackage{psfrag}
\begin{document}
\date{\today}
\title{Relationship between the thermopower and entropy of strongly
          correlated electron systems}
\author{V. Zlati\'c$^{1,2}$, R. Monnier$^3$, J. Freericks$^4$, K. W. Becker$^2$}
\affiliation{$^1$Institute of Physics, Bijeni\v cka c. 46, 10001 Zagreb, Croatia (permanent address)}
\affiliation{$^2$Institut f\"ur Theoretische Physik, 
Technische Universit\"at Dresden,  01062 Dresden, Germany}
\affiliation{$^3$ETH H\"onggerberg, Laboratorium f\"ur Festk\"orperphysik, 8093 Z\"urich, Switzerland}
\affiliation{$^4$Georgetown University, Washington D.C., USA}

\begin{abstract}
A number of recent experiments report the low-temperature  
thermopower $\alpha$ and specific heat coefficients $\gamma=C_V/T$ 
of strongly correlated electron systems. 
Describing the charge and heat transport in a thermoelectric by transport 
equations, and assuming that the charge current and the heat current 
densities are proportional to the number density of the charge carriers, 
we obtain a simple mean-field relationship between 
$\alpha$ and the entropy density $\cal S$ of the charge carriers. 
We discuss corrections to this mean-field formula and use results obtained for 
the periodic Anderson and the Falicov-Kimball models
to explain the concentration (chemical pressure) and temperature dependence 
of $\alpha/\gamma T$ in EuCu$_2$(Ge$_{1-x}$Si$_x$)$_2$, CePt$_{1-x}$Ni$_x$, 
and YbIn$_{1-x}$Ag${_x}$Cu$_4$ intermetallic compounds. 
We also show, using the 'poor man's mapping'  which approximates the periodic Anderson  
lattice  by the single impurity Anderson model, that the seemingly complicated behavior 
of $\alpha(T)$ can be explained in simple terms and that the temperature dependence 
of  $\alpha(T)$ at each doping level is consistent  with the magnetic character of 4{\it f} ions. 
\end{abstract}
\maketitle
\section{ {Introduction} \label{Introduction}}  
Several recent papers\cite{sakurai.93,behnia.04,sakurai.05}
report on a  correlation between the low-temperature  thermopower $\alpha$ 
and the specific heat coefficient $\gamma=C_V/T$ for heavy Fermions 
and valence fluctuating compounds with Ce, Eu, Yb and U ions.   
The data show that the zero-temperature limit of the ratio 
$\alpha(T)/C_V(T)=\alpha/\gamma T$ in most systems is about the same, 
although the absolute values of  $\gamma$ and  $ \alpha/T$ vary by orders of magnitude
\cite{sakurai.93,behnia.04,sakurai.05,sakurai.02,espeso.94,fukuda.03,hossain.04,ocko.01,ocko.04,benz.99,loehenysen.94}. 
At the moment it is not clear if the deviations from universality arise because of insufficient 
accuracy of the data, or because the `universal law' is only approximately valid. 
The difficulty is that neither $\alpha(T)$ nor $C_V(T)$ are linear in the lowest available 
temperature intervals and to obtain the linear coefficients one has to estimate the functional 
form of $\alpha(T)$ and $C_V(T)$ before taking the zero-temperature limit. 
The temperature dependences of $\alpha(T)$ and $C_V(T)$ in various systems 
might have a different physical origin, and comparing the data on $\alpha(T)/\gamma T$  
without a detailed knowledge of the underlying dynamics might lead to erroneous 
conclusions. 
Furthermore, the definition of the `zero-temperature limit'  can differ considerably 
for systems with vastly different characteristic temperatures  and at present it is not 
clear what the error bars are for the $\alpha/\gamma T$ data. 
We believe, the universality of  the $\alpha/\gamma T$ ratio could be tested 
most directly by performing a pressure experiment which transforms 
Ce- or Eu-based heavy Fermion materials into  valence fluctuators, 
and Yb-based valence fluctuators into  heavy Fermion materials. 
Pressure can not only change the characteristic temperature of 
a given compound by several orders of magnitude\cite{wilhelm.02,wilhelm.05}  
and, with that, change the nature of the ground state, but it has a dramatic effect 
on the overall shape of the thermopower. 
Although the low-temperature pressure experiments are less accurate than the 
ambient pressure ones, the evolution of  $\alpha(T)$ or $C_V(T)$ with pressure could 
be followed and the universality of $\alpha/\gamma T$ studied in a systematic way.  
Experiments which provide the low-temperature thermopower and 
the specific heat on the same sample at various pressures are yet to be
performed but some data are available on the doping dependence 
(chemical pressure) of $\alpha$ and $\gamma$. 
For example, doping alters $\alpha$ and $\gamma$  of 
EuCu$_2$(Ge$_{1-x}$Si$_x$)$_2$, CePt$_{1-x}$Ni$_x$ and 
YbIn$_{1-x}$Ag${_x}$Cu$_4$  intermetallics by more than one order of magnitude
\cite{fukuda.03,hossain.04,sakurai.02,espeso.94,ocko.01,ocko.04}.  
while barely affecting the low-temperature ratio $\alpha/\gamma T$.  
We explain this behavior as a chemical pressure effect. 

A simple relation between $\alpha$  and ${\gamma T}$ is obtained for 
a free electron gas and for non-interacting electrons on a lattice. 
The solution of the Boltzmann equation gives, under fairly general 
conditions, the result\cite{joffe,ashcroft&mermin}  
$\alpha(T)\simeq{\cal S}(T)/ne= C_V(T)/ne$, 
where ${\cal S}$ is the entropy density, $n$ the number density 
of the charge carriers and the second equality holds because the specific
heat is linear with $T$ at low $T$. But non-interacting electrons fail to 
describe the properties of the compounds mentioned above and the question arises 
to what extent does the `universal  relation'  between  $\alpha$ and ${\cal S}$
(or $C_V$) hold for strongly correlated electrons?

On a macroscopic level, an analysis of the charge and heat transport of a 
thermoelectric in terms of transport equations yields the same relationship 
between $\alpha$ and $\cal S$  as the free electron model. 
This derivation assumes that under isothermal conditions the 
expectation value of the charge current 
density is proportional to the expectation value of the
heat current density, and we expect it to be valid 
for Fermi liquids (FL); of course, this is neglecting other sources of
heat transport, such as phonon contributions (but those do not enter into
isothermal heat transport unless there are phonon drag effects). 
In the case of the non-FL compounds, where $C_V(T)/T$ and $\alpha(T)/T$ 
are temperature-dependent,  instead of $\alpha/T\gamma$ one should consider  
the ratio $ \alpha/{\cal S}$. 

The zero-temperature $\alpha/\gamma T$-ratio has been discussed recently 
by Miyake and Kohno\cite{miyake.05}. They calculate the dynamical conductivity  
of the periodic Anderson model in the quasiparticle (QP) approximation and 
obtain  the Seebeck coefficient from the canonical formula which expresses 
$\alpha(T)$ as the ratio of two transport integrals\cite{mahan.81}. 
The universal  FL  behavior is derived by assuming that in addition 
to the renormalization effects due to the Coulomb interaction between {\it f}-electrons,   
the charge and heat transport are affected by the scattering of quasiparticles 
off residual impurities. This scattering dominates the transport relaxation time 
and gives rise to a finite residual resistivity $\rho_0$. 
In the dilute limit, the impurity concentration drops out of the ratio of 
transport integrals and does not appear in the expression for the Seebeck coefficient. 

The exact many-body transport coefficients of the periodic spin-1/2 Anderson 
model have recently been obtained at finite temperatures by 
Grenzebach et al.\cite{grenzebach.06} 
using the dynamical mean field theory (DMFT) which maps the lattice problem 
onto an auxiliary single impurity model with a self consistency
condition. The electron relaxation is solely due to the 
Coulomb repulsion between {\it f}-electrons and the model describes 
both heavy Fermions and valence fluctuators with vanishing $\rho_0$. 
The solution of the auxiliary impurity problem is obtained by the numerical 
renormalization group (NRG) method, which provides accurate results for the static properties 
at arbitrary temperature and for the dynamical properties above the FL regime. 
However, the zero-temperature limit of the transport relaxation time
is difficult 
to obtain by the DMFT+NRG method (due to numerical issues associated with
the self-consistency), and the validity of the FL laws for the 
transport coefficients of strongly correlated electrons has not been
clearly established.  

In this contribution, we discuss the  zero-temperature $\alpha/\gamma T$-ratio 
for the periodic Anderson model and the Falicov-Kimball model, which describe 
heavy Fermions and valence fluctuators with crystal field (CF)  split 4{\it f}-states. 
In this limit  the excited CF states are not occupied and we calculate the slope of 
$\alpha(T)$ using the DMFT solution of an effective $\cal M$-fold degenerate model.    
The result agrees with that of Miyake and Kohno\cite{miyake.05} even though 
we consider a periodic model with vanishing  $\rho_0$
and have a positive coefficient for the $T^2$  term of the electrical resistance. 
We  also analyze the influence of CF splittings on the thermopower of the 
Anderson model above the FL regime. At  present, the dynamical properties of 
such a model cannot be obtained by the exact DMFT mapping and we find the 
solution using an approximate `poor man's mapping'. 
That is, we assume that the conduction electrons scatter incoherently off the 
4{\it f}-ions,  relate the transport relaxation time to the single-ion T-matrix,   
and solve the scattering problem by the non-crossing approximation (NCA), 
which can properly treat the highly asymmetric limit of the Anderson model 
and infinite Coulomb repulsion. 
The results obtained in such a way explain the concentration (chemical pressure) 
dependence of  $\lim_{T\to 0} \alpha/\gamma T$  and the evolution of $\alpha(T)$ 
in  EuCu$_2$(Ge$_{1-x}$Si$_x$)$_2$\cite{sakurai.05,fukuda.03,hossain.04} 
CePt$_{1-x}$Ni$_x$\cite{sakurai.02},  and 
YbIn$_{1-x}$Ag${_x}$Cu$_4$\cite{ocko.01,ocko.04,sarrao.96,cornelius.97}. 
The thermoelectric properties of these  systems exhibit all the typical features 
observed in heavy Fermions and valence fluctuators\cite{jaccard.96,zlatic.05}. 
Our calculations show that the temperature dependence of $\alpha(T)$ 
at each doping level is consistent with the character of the ground state inferred 
from the initial thermopower slope. 

The rest of this contribution is organized as follows. In Section II,  we introduce the 
macroscopic transport equations, discuss the Seebeck and Peltier experiments, 
and find the relationship between the thermopower and the entropy. 
In Section III.1 we calculate the Seebeck coefficient of the periodic Anderson 
model in the FL regime using the DMFT mapping. In Section III.2 we calculate  the 
finite-temperature behavior using the `poor man's mapping'.  
In Section  III.3 we discuss the thermoelectric properties of the
Falicov-Kimball model using the DMFT approach. 
In Section IV,  we use these results to discuss the experimental data on the 
intermetallic compounds mentioned above. 
Our conclusions are presented in Section V.

\section{ {Transport equations} \label{Transport} }
We consider the macroscopic  charge  and  energy currents which are given
by the statistical averages ${\bf J} = {\mathrm Tr} \{ \rho_\phi \hat{\bf j}\}$  
and ${\bf J}_{E}^{\phi} = {\mathrm Tr} \{ \rho_\phi \hat{\bf j}_{E}^{\phi} \}$,  
where $\rho_\phi$ is the density matrix, $\hat{\bf j}$ the charge current  density 
and $\hat{\bf j}_{E}^{\phi}$  the energy current density operators for a system 
of charged particles in the presence of an external scalar potential $\phi$. 
These operators are are obtained by commuting the Hamiltonian with the 
charge and energy polarization operators\cite{mahan.81} respectively,
which are chosen 
in such a way that the macroscopic currents satisfy the appropriate continuity
 equations\cite{callen.48}. 
Assuming that the external potential couples to the charge density, a
direct calculation shows\cite{luttinger.64}  that $\hat{\bf j}$ is field independent 
and $\hat{\bf j}_{E}^{\phi} = \hat{\bf j}_{E} + \phi\hat{\bf j}$, where $\hat{\bf  j}_{E} $ 
is the energy current density defined by the field-free Hamiltonian. 
The macroscopic energy current ${\bf J}_{E}= {\mathrm Tr} \{ \rho_\phi \hat{\bf j}_{E} \}$
does not satisfy the continuity equation in the presence of the external potential 
 but is easily calculated by a perturbative expansion. 
The gradient expansion of  $\rho_\phi$ produces 
linearized equations\cite{luttinger_note}, 
${\bf J}=L_{11} {\bf x}_c + L_{12} {\bf x}_E$ 
and ${\bf J}_E = {\bf J}_{E_\phi} - \phi{\bf J}=L_{21} {\bf x}_c + L_{22} {\bf  x}_E$, 
where the generalized forces are given by 
${\bf x}_c= -\nabla\phi - T\nabla(\mu/eT) $ and ${\bf x}_E= - {\nabla T}/{T} $,  
$\mu$ is the chemical potential and $e=-|e|$ the electron charge.  
The (linear-response) expansion coefficients are given by the correlation functions, 
\begin{equation} 
             \label{luttinger}
L_{ij}^{\alpha\beta}=\lim_{s\to 0} \frac 1 V 
\int_0^\infty dt e^{-st}
\int_0^\beta  d\beta 
\langle\hat {\bf j}_{i}^\alpha(-t-i\beta)\hat {\bf j}_{j}^\beta(0)\rangle_0  ,  
\end{equation} 
where $\langle \hat{\bf j}_{i}^\alpha \hat{\bf j}_{j}^\beta\rangle_0
= {\mathrm Tr} \{ \rho \; \hat{\bf j}_{i}^\alpha \hat{\bf j}_{j}^\beta \}$ 
denotes the statistical average in the absence of the external potential and 
$\hat{\bf j}_1^\alpha$ and $ \hat{\bf j}_2^\beta$  denote the $q=0$ Fourier components of 
$\hat {\bf j}^\alpha(x)$ and $\hat{\bf j}_E^\beta(x)$, respectively ($\alpha, \, \beta$ denote the 
coordinate axes).  
In what follows, we assume a homogenous and isotropic conductor 
in the absence of  a magnetic field and consider only a single Cartesian component 
of $\hat{\bf j}$ and $\hat{\bf j}_E$ (this is appropriate for cubic systems).

Thermoelectric effects are usually described in terms of the heat current, 
hence we transform the linearized equations from ${\bf J}$ and ${\bf J}_E$ 
to ${\bf J}$ and ${\bf J}_Q={\bf J}_E-(\mu/e){\bf J}$ to 
yield\cite{landau,mahan.81}
\begin{eqnarray} 
                   \label{*}
{\bf J}&=&-\sigma\nabla\phi- \sigma\alpha \nabla {T}, \\
                   \label{**}
{\bf J}_Q&=&\alpha T {\bf J} - \kappa {\nabla} T,   
\end{eqnarray} 
where $\sigma=L_{11}$, $\alpha T = (L_{12}/L_{11}-\mu/e)$, and 
$\kappa T=(L_{22}-L_{12}^2/L_{11})$. 
A simple analysis shows that $\sigma(T)$, $ \alpha(T) $, and $\kappa(T)$ are 
the isothermal electrical conductivity,  the Seebeck coefficient and the 
thermal conductivity, respectively\cite{domenicali.54,landau}. 
The Onsager relation\cite{Onsager.31} gives $\alpha T = \Pi$, 
where $\Pi$ is the Peltier coefficient. 

The stationary temperature distribution across the sample 
is obtained from the total energy current in a field
${\bf J}_{E}^{\phi} = {\bf J}_{Q} + (\phi+\mu/e) {\bf J} $
which satisfies the energy continuity equation\cite{callen.48} 
$ {\mbox{div}} {\bf J}_{E}^{\phi} =0$
and 
leads to the Domenicali equation\cite{domenicali.54}, 
\begin{equation} 
                             \label{domenicalli}
\dot{ {E}_{\phi} }={\mbox{div}}(\kappa\nabla T)+ \frac{ {\bf J}^2 }{\sigma}-T{\bf J} \cdot{\nabla}\alpha =0  .
\end{equation} 
The Joule term $(J^2/\sigma)$ in Eq.~\eqref{domenicalli} is crucial for the total
energy balance and 
for establishing the correct steady-state temperature distribution. 

The solution of  Eqs.~\eqref{*}--\eqref{domenicalli},  with appropriate boundary 
conditions, completely specifies the thermoelectric response of the system. 
The above procedure connects the theoretical model of a given material 
with the phenomenological transport coefficients and could be used, in principle, 
to explain the experimentally established  relationship between the thermopower  
and the specific heat. 
However, for general many body systems such a program cannot be completed 
and, at  first sight, it is not obvious that the transport coefficients, which depend 
on the dynamical properties of the system, are simply related to the thermodynamic 
quantities, which depend only on the static properties\cite{joffe}.
For example, the heat current considered in this many-body theory is
just the electronic contribution to the heat transport, hence we only
expect it to relate to the appropriate electronic contributions to the 
entropy.
On a macroscopic level, the relationship  between $\alpha$ and $\cal S$ can 
be obtained by solving Eqs.~(\ref{*}) and (\ref{**}) once with the boundary 
conditions corresponding to the measurement of the Seebeck coefficient and 
once with those appropriate for the measurement of the Peltier coefficient.

In the Seebeck setup (an open circuit without net charge current), 
the thermoelectric voltage is induced by the heat flow due to the temperature gradient. 
The Seebeck voltage appears because the charged particles diffuse from the hot 
to the cold end and the imbalance of charge gives rise to a potential gradient 
across the sample. The Seebeck coefficient is obtained from the ratio 
${\Delta V}/{\Delta T}$,  where $\Delta V=-\int_{0}^{a} dx \ {\nabla\phi ({\bf x)}}$  
is the voltage change and $\Delta T=\int_{0}^{a}  dx \  {\nabla T({\bf x)}}$ 
the temperature drop between the end-points of a sample of length $a$.  
For constant $\alpha(x)$, Eq.~(\ref{*}) gives ${\Delta V}= \alpha\Delta T$.  
In a stationary state, a quasiparticle picture 
says that the electrical energy required to transfer $n$ electrons from 
the hot end to the cold end against the voltage ${\Delta V}$ 
is balanced by the change in thermal energy (that is, the heat).  
Neglecting the shift of the quantum states due to the external potential, 
we approximate $ne{\Delta V}\simeq {\cal S}_n  \Delta T  $, 
where  $n$ is the particle density and ${\cal S}_n$ the entropy density 
of the charge carriers. This gives the approximate relationship between 
the Seebeck coefficient and the entropy as
\begin{equation} 
                             \label{alpha-S}
\alpha(T) =\frac{{\cal S}_n(T)}{en}. 
\end{equation} 

In the Peltier setup, a constant electrical current passing through a junction of two 
different thermoelectrics gives rise to an additional heat current emanating at 
the interface of the junction. 
The Peltier heat appears  because the excitation spectra 
on the two sides of the interface are different, 
so that the charge transfer produces an entropy change. 
(The entropy of $n$ particles is determined by the structure of the energy 
levels over which the current carriers are distributed.) 
In a stationary state,  the normal component of currents and temperature  are 
continuous  across the junction but $\nabla T$ and the transport coefficients 
are discontinuous. Integrating Eq.~(\ref{domenicalli}) over a thin volume 
element containing  the interface gives\cite{landau}, 
\begin{equation} 
                                             \label{eq:Peltier} 
-\kappa_s \nabla T|_s - (-\kappa_l \nabla T|_l)
= 
 -{\bf J} T (\alpha_s - \alpha_l) ,
\end{equation}  
where we used Gauss theorem and neglected the Joule heat. 
$\kappa_s$, $\nabla T|_s$, $\alpha_s$ and $\kappa_l$, $ \nabla T|_l$, $\alpha_l$
denote the thermal conductivity, temperature gradient and the Seebeck coefficient 
of the `sample' and `leads' in the vicinity of the interface. 
Eq.~(\ref{eq:Peltier}) shows that the heat brought to and taken from the surface by  
the thermal conductivity differs by $\Pi_{sl} {\bf J}$, where $\Pi_{sl}$ is the relative Peltier 
coefficient of the two materials,  $\Pi_{sl}=T[\alpha_s(T) - \alpha_l(T)]$.
In other words, the junction generates an additional heat current $\Pi_{sl} {\bf J}$ 
which maintains the stationary state by absorbing (or releasing) 
heat from the environment. 
Depending on the sign of ${\bf J}$, the Peltier heat is either 
absorbed or released,  i.e., the entropy change due to the Peltier effect is reversible. 

Under stationary isothermal conditions, and for currents flowing in the x-direction,
Onsager's relation 
gives 
\begin{equation} 
                                             \label{eq:seebeck}
\alpha(T)=\frac{J_Q}{TJ} , 
\end{equation}
where $J_Q$ is the Peltier heat current. 
Neglecting  the Joule heat and assuming that the stationary state is maintained by
a heat source at one end and a sink at the other end of the sample, 
we can simplify Eq.~\eqref{eq:seebeck}  using  $\mbox{div }{\bf J}=0$ 
and $\mbox{div }{\bf J}_Q=0$.  For such divergence-free currents, 
we assume that the charge and heat flow uniformly 
with a drift velocity ${\bf v}$ and write ${\bf J}=ne {\bf v}$ and ${\bf J}_Q=  Q_n{\bf v}$, 
where $Q_n=\alpha T ne $ is the Peltier heat generated at the lead-sample
interface and transported by the current in the lead to the sink. 
Defining the reversible thermoelectric entropy density as ${\cal S}_n(T)=Q_n/T$, 
we reduce Eq.~\eqref{eq:seebeck} to Eq.~\eqref{alpha-S}. 
Finally, multiplying both sides of Eq.~(\ref{alpha-S}) by  $N_{A}e$, where $N_A$ is 
Avogadro's constant,  
 and 
dividing by the molar entropy
${\cal S}_{N}(T)={\cal S}_{n}(T)\Omega$, where $\Omega$ is the molar volume
of the material under study, we obtain a dimensionless parameter
\begin{equation} 
                            \label{gg}
q = N_{A} \frac{e\alpha(T)}{{\cal S}_{N}(T)} = (\frac{N}{N_A})^{-1},
\end{equation} 
which characterizes the thermoelectric material in terms of an effective charge carrier 
concentration per formula unit  (or the Fermi volume $V_F$ of the charge carriers).  
For Fermi liquids, ${\cal S}(T)= \gamma T$ at low temperature and we obtain
\begin{equation} 
                            \label{g}
q= N_{A}e \frac{\alpha(T)}{\gamma T} , 
\end{equation} 
which is  the quantity defined by Behnia, {\it et al.}\cite{behnia.04}. 
Throughout this paper, $\alpha$ is expressed in [$\mu$V/K],  
$C_V$ and ${\cal S}$ in [J/(K~mol)], and 
the Faraday number is $N_A e=9.6\times 10^{4}$C/mol. 

Before proceeding, we should comment on the validity of the above approximations. 
As mentioned already, the entropy of the charge carriers in the steady state 
which characterizes the Seebeck setup is not the same as the equilibrium entropy, 
because the steady-state potential is different at the hot and the cold end.  
As regards the Peltier setup, the average value of the current density operator 
is not simply proportional to the particle density, and the definition used in 
Eq.~\eqref{gg} neglects all the operator products which lead to  higher-order 
powers  in the particle density.  This amounts to describing the low-energy excitations 
of the system by quasiparticles and approximating the many-body interactions by  
self-consistent fields.

Furthermore, we should take into account that the entropy $ {\cal S}_{n}$ in 
Eq.~(\ref{alpha-S}) or  $ {\cal S}_{N}$ in Eq.~\eqref{gg} is not the 
full entropy ${\cal S}_{}$ of the system, but only the 
entropy of the charge carriers appearing in the transport equations.
For example, the total entropy ${\cal S}_{}$  might have contributions ${\cal S}_{M}$ coming 
from additional degrees of freedom, like localized paramagnetic states, 
magnons or phonons, which do not participate in the charge transport 
and are only weakly coupled to the charge carrying modes.  
Assuming ${\cal S}_{}= {\cal S}_{N} + {\cal S}_{M}$ but neglecting the contribution of these 
additional degrees of freedom to the charge transport, we get the experimentally 
determined quantity 
\begin{equation} 
                            \label{ggg}
\tilde q={N_{A}}\frac{e\alpha_{}(T)}{{\cal S}_{}(T)}
= \frac{N_{A}}{N} 
\frac{1}{ 1 + {\cal S}^{}_{M}(T)/{\cal S}^{}_{N}(T) } ,
\end{equation} 
which could be much  reduced with respect to $ q= {N_{A}}/{N}$ given by 
Eq.~(\ref{gg}). 
The experimental values $\tilde q$ depend not only on the concentration of carriers 
but also on temperature, and to get the universal ratio one might need a very 
low temperature, where $ {\cal S}_{M}\ll {\cal S}_{N}$.
This behavior is similar to deviations of the Wiedemann-Franz law from ideal
metallic behavior whenever the phonon contribution to the heat current is 
substantial---in order for Wiedemann-Franz to hold, we need to have the 
phonon contribution to the thermal conductivity be much smaller than 
the electronic contribution.

The Seebeck coefficient  appearing in  Eqs.~(\ref{alpha-S})---(\ref{ggg}) should 
also be treated with care. If there are several conductivity channels, the total thermopower is 
a weighted sum of all the components $\sigma \alpha =\sum_j \sigma_j \alpha_j$, 
where $\sigma =\sum_j \sigma_j $, and there might be some cancellations 
in the thermopower sum. 
But ${\cal S}$ has different vertex corrections and the specific heat is not affected 
by these cancellations. 
Similarly, if there are several scattering channels for conduction electrons, 
vertex corrections give rise to  interferences which affect the thermopower 
(like in the Friedel phase shift formula\cite{costi.94}). 
Even if we neglect interference effects and use the Nordheim-Gorter rule\cite{barnard.72} 
($\rho \alpha =\sum_j \rho_j \alpha_j$, where $\rho =\sum_j \rho_j $ and 
$\rho_j$ and $\alpha_j$ are the resistivity and thermopower  due to the j-th  scattering channel),  
the $\alpha_j$-terms in the  weighted sum might have different signs and cancel.  
Thus, unless one of the channels dominates, $\tilde q$ is non-universal and 
temperature-dependent, and the interpretation becomes difficult. 
We also  remark that the heat conductivity of magnons and phonons can give 
rise to  phonon-drag  and spin-drag contributions to $\alpha_{}(T)$, 
which are not included in Eqs.~(\ref{gg})  or (\ref{ggg}).
However, at low temperatures, those contributions are expected to be small.  

For systems with a non-linear $C_V(T)$, like Yb and Ce compounds close to a
quantum critical point\cite{benz.99,loehenysen.94}, 
the experimental data cannot be analyzed in terms of Eq.~(\ref{g}), 
even at low temperatures,  but one could use Eq.~(\ref{gg}) instead. 
However,  if the single-particle states are not well defined and thermal transport is due 
to collective excitations rather than quasiparticles, like in a Luttinger liquid\cite{kane.96}, 
the significance of $N/N_{A}$ as the number of the current carrying particles 
per formula unit, is no longer obvious. 

\section{Model calculation of the Seebeck coefficient\label{models}}
Considering the limitations and uncertainties mentioned above, it is somewhat 
surprising that in many correlated systems the low-temperature ratio of the thermopower 
and the specific heat comes quite close to the universal value given by Eq.~\eqref{g}.  
In what follows, we show that the universal law of Sakurai\cite{sakurai.93}  
and Behnia\cite{behnia.04}   holds for the periodic Anderson model with on-site 
hybridization  and for the Falicov-Kimball model. We also show that these models 
explain the full temperature dependence of the Seebeck coefficient observed 
in the intermetallic compounds with Ce, Eu and Yb ions. 
The charge current operator in both models  is 
\begin{equation}
\hat{\bf j}=e\sum_{{\bf k}\sigma} {\bf v}_{\bf k}c^\dagger_{{\bf k}\sigma}c^{}_{{\bf k}\sigma} , 
                         \label{eq: charge_current2}
\end{equation}
where $\sigma$  labels the symmetry channels (irreducible representations 
to which the conduction electrons belong)  and 
${\bf v}_{\bf k}=\nabla\epsilon({\bf k})$ is the velocity  of unperturbed conduction electrons. 
Calculating the heat current density operators for a constant hybridization in $\bf k$-space  
we verify explicitly the Jonson-Mahan theorem \cite{mahan.98} and find 
for each symmetry channel the static conductivity  satisfies
\begin{equation}
                           \label{sigma}                           
\sigma(T) 
= 
 \int d\omega \left ( -\frac{df}{d\omega}\right )\Lambda(\omega,T)
\end{equation}
and the thermopower satisfies
\begin{equation} 
                           \label{alpha}
\alpha(T)=-
\frac{1}{|e|T}
\frac{ \int d\omega \left ( -\frac{df}{d\omega}\right )\omega\Lambda(\omega,T)}{\int d\omega \left ( -\frac{df}{d\omega}\right )\Lambda(\omega,T)}. 
\end {equation}  
The excitation energy $\omega$ is measured with respect to $\mu$, 
$f(\omega)=1/[1+\exp(\beta\omega)]$ is the Fermi-Dirac distribution function and 
$\Lambda(\omega,T)$ is the charge current - charge current correlation 
function\cite{mahan.81} (our result differs from Mahan's by an additional factor 
of $e^2$ from the charge current operators). 
In the low-temperature FL limit, the charge current - charge current correlation 
function is approximately found from the reducible vertex function by
\begin{eqnarray} 
                            \label{Lambda}
\Lambda(\omega,T)=
\frac{e^2}{V} 
&\sum_{{\bf k}\sigma}&
 {\bf v_{\bf k}}^2 G_c^\sigma({\bf k},\omega+i\delta)G_c^\sigma({\bf k},\omega-i\delta)\nonumber\\
&\times&\gamma^\sigma({\bf k},\omega+i\delta,\omega-i\delta)  . 
\end {eqnarray} 
Here $G_c^\sigma({\bf k},\omega\pm i\delta)$ are the momentum and 
energy-dependent retarded and advanced Green's function of the conduction electrons 
and $\gamma^\sigma({\bf k},\omega+i\delta,\omega-i\delta)$ 
is the analytic continuation from the imaginary axis into the complex plane 
of the (reducible) scalar vertex function  $\gamma^\sigma({\bf k},i\omega_n)$,  which is defined 
by the diagrammatic expansion of the current-current correlation 
function\cite{mahan.81}.   
The details of the calculations are model-dependent and, in what follows, 
we consider separately the periodic Anderson model and the Falicov-Kimball model.
One can also determine the charge current - charge current correlation function
exactly within DMFT, where the vertex corrections vanish, and
\begin{equation}
\Lambda(\omega,T)=\frac{e^2}{V}\sum_{{\bf k}\sigma}{\bf v}_{\bf k}^2
\left [ {\rm Im} G^\sigma_c({\bf k},\omega+i\delta)\right ]^2.
                            \label{eq: corr_fcn_dmft}
\end{equation}
This form is particularly useful if  $G^\sigma_c({\bf k},\omega+i\delta)$ 
is known for all ${\bf k}$ and $\omega$ points; we use it for evaluating the
Falicov-Kimball model\cite{freericks.03a,freericks.01}, 
since it does not have FL behavior in general. 
 
{\subsubsection{Periodic Anderson model - the low-temperature DMFT solution}\label{anderson model}}
The periodic Anderson model is defined by the Hamiltonian 
\begin{equation}
                               \label{Hamiltonian}
H_{A}=H_{c}+H_{f}+H_{cf},
\end{equation}
where $H_{c}$ describes the conduction electrons hopping on the lattice,  
$H_{f}$ describes the 4{\it f}  states localized at each lattice site, and 
$H_{cf}$ describes the hybridization of electrons between 4{\it f} and conduction states. 
The total number of conduction electrons and {\it f}-electrons  per site is $n_c$ 
and $n_f$, respectively. 
We consider the model with an infinitely strong Coulomb repulsion between 
{\it f}-electrons (or  {\it f}-holes)  which does not allow two {\it f}-electrons
to occupy the same 
lattice site. That is, we strictly enforce the constraint $n_f\leq 1$ (or $n_f^h\leq 1$), and 
also choose $n_c\leq 1$. The total number of electrons $n=n_c+n_f$ is conserved. 
The unrenormalized density of  states (DOS) of the conduction electrons 
in each channel is   
${\cal N}_c^0(\epsilon)=\sum_{\bf k} \delta(\epsilon-\epsilon_{\bf k}^{})$, where 
$\epsilon_{\bf k}$ is the conduction-electron dispersion. 
We assume ${\cal N}_c^0(\epsilon)$ to be a symmetric, slowly varying function 
of half-width $D$, which is the same in all channels,  
and measure all the energies, except $\omega$, with respect to its  center. 
In the case of heavy Fermions and valence fluctuators with Ce ions, 
the non-magnetic state is represented by an empty {\it f}-shell 
and the magnetic state by the CF levels. 
We consider ${M}-1$ excited CF levels separated from the CF ground state 
by energies $\Delta_i\ll |E_f|$, where $i=1,\ldots,{ M} -1$. 
Each CF state belongs to a given irreducible representation of the point group 
$\Gamma_i$ and its degeneracy is ${\cal L}_i$, such that $ {\cal L}=\sum_{i} {\cal L}_i $.  
The spectral functions of the unrenormalized {\it f}-states are given by a set of delta 
functions at $E_f^{0}$ and $E_f^{i}=E_f^{0}+\Delta_i$.
The mixing matrix element $V_i$ connects the conduction and {\it f}- states 
belonging to the same irrep and, for simplicity,
we consider the case $V_i= V$ for all channels, i. e. the hybridization is
characterized by the parameter $\Gamma=\pi V^2\mathcal{N}_c^0(0)$. 
The conduction and $f$-states have a common chemical potential $\mu$.
The properties of the model depend in an essential way on the coupling 
constant $g= \Gamma/\pi |E_f-\mu|$,  where 
 $E_f=\sum_{i} {\cal L}_i E_f^i/\cal L$ 
and on  the CF splittings.  
For ${\cal L} > 2$ and strong correlation between the  {\it f}-electrons, 
the condition  $n_f \leq 1$ makes the model extremely asymmetric; 
$n_f\simeq 1$ can only be reached for very small $g$. 
In such a model, an increase in $g$ gives rise to a monotonic reduction 
of $n_f$, which assumes a characteristic value at each fixed point. 
An application of pressure or chemical pressure (doping) to Ce systems 
is modeled by an increase of the bare coupling $g$, which decreases $n_f$. 
In the case of Europium the non-magnetic 4{\it f}$^{~6}$ state is a Hund's singlet 
and the magnetic state is a degenerate 4{\it f}$^{~7}$ Hund's octet with full 
rotational invariance.  The Eu ion fluctuates between the two configurations 
by exchanging a single electron with the conduction band and we model 
the pressure effects in the same way as for Ce. 
In the case of Ytterbium, the  non-magnetic state is the full-shell 4{\it f}$^{~14}$ 
configuration and the magnetic one is the 4{\it f}$^{~13}$ configuration, 
which can be split by the CF. Here, pressure or chemical pressure reduce $g$ 
and enhance the number of {\it f}-holes. 

We first consider the CF and spin degenerate case 
($\Delta_i=0$ for all $\mathcal{L}$ $f$-states) and 
treat the correlations by  the DMFT which is exact in the limit 
of infinite dimensions\cite{metzner_vollhardt_1989,brandt1989,kotliar_review}.
To simplify the notation we drop the spin-label in this subsection. 
Using the equations of motion for the imaginary time Green's functions,  
making the Fourier transform to Matsubara frequencies
and analytically continuing into the complex 
energy plane, we obtain the Dyson equations for the {\it c}- and {\it f}-electrons Green's 
functions, 
\begin{equation} 
                            \label{G_c} 
 G_c({\bf k},z)
 =
\frac{z -(E_f-\mu)-\Sigma_f({\bf k},z)} 
{[z -(\epsilon_{\bf k}-\mu)][z -(E_f-\mu)-\Sigma_f({\bf k},z)]-V^2} ,
\end {equation} 
and
\begin{equation} 
                            \label{G_f}
G_f({\bf k},z)
=
 \frac{z -(\epsilon_{\bf k}-\mu)} 
{[z -(\epsilon_{\bf k}-\mu)][z -(E_f-\mu) -\Sigma_f({\bf k},z)]-V^2} , 
\end{equation}  
where $z$ satisfies
$z=\omega+ i\delta$ and $\Sigma_f(k,z)$ is the  {\it f}-electron self energy
(the self energy and Green's functions are identical for each of the
$\mathcal{L}$ different $f$-states). 
The retarded (advanced) Green's functions are defined by the above expressions 
for $z$ in the upper (lower) part of the complex plane and are denoted on the 
real axis by $\omega^\pm= \lim_{\delta\to 0} (\omega\pm i\delta)$. 
The self energy of the conduction electrons is 
\begin{equation} 
                            \label{sigma_c}
\Sigma_c({\bf k},z)
=
\frac{V^2}{z -(E_f-\mu)-\Sigma_f({\bf k},z)} .
\end{equation}
In  DMFT, the self-energies are momentum independent:  
$\Sigma_f({\bf k},z)=\Sigma_f(z)$ and   $\Sigma_c({\bf k},z)=\Sigma_c(z)$, 
and the model is solved by mapping the local  Green's function,  
$G_f(z)=\sum_{{\bf k}}G_f({\bf k},z)$, onto the Green's function of an auxiliary 
single impurity Anderson model, 
\begin{equation} 
                                  \label{f-DMFT}
G_f(z) 
=
 \frac{1}{z - (E_f - \mu) - \Delta(z) - \Sigma_f(z) } , 
\end {equation} 
where  $\Delta(z)$ is determined by the requirement 
that  Eqs.~\eqref{G_c} - \eqref{f-DMFT} are  self-consistently solved and that
$\sum_{\bf k}G_f({\bf k},z)$ is equal to $G_f(z)$ in Eq.~\eqref{f-DMFT}. 
The irreducible self energy of the lattice and the impurity are given 
by the same functional, 
$\Sigma_f[\sum_{{\bf k}}G_f({\bf k},\omega)]=\Sigma_f[G_f(\omega)]$, 
and the self-consistency requires that  the auxiliary spectral function, 
$A(\omega)= -{\mathrm Im} \  G_f(\omega^+)/\pi $ coincides with 
the  local {\it f}-DOS of the lattice,  
${\cal N}_f(\omega)=  \sum_{\bf k} A_f({\bf k},\omega)$, 
where 
\begin{equation} 
                            \label{A_f_1}
A_f({\bf k},\omega)
=
 - \frac{1}{\pi} {\mathrm Im} \ G_f({\bf k},\omega^+).
\end{equation}
The impurity model generated by the DMFT describes a single {\it f}-electron 
distributed over $\cal L$-fold  degenerate spin or CF states and 
the double occupancy of these states is explicitly forbidden, $n_f\leq 1$. 
The energy of the effective impurity {\it f}-state is $E_f$ and the coupling to the 
auxiliary conduction band is described the hybridization function $\Delta(\omega^+) $. 
Since $n_f$ is the same on the impurity  and on the lattice,  
for large $\cal L$  the impurity model is highly asymmetric.  
For $\omega\simeq 0$, we assume that the $\omega$-dependence of $\Delta(\omega)$ is much 
slower than of ${\mathrm Im}\ \Sigma_f(\omega^+,T)$ and 
approximate, $\Delta(\omega)\simeq \Delta(0)=i\Delta_0$.  
The positive-definiteness of $ {\cal N}_f(\omega)$ requires    $\Delta_0 < 0$. 

The DMFT simplifies the calculation of the current-current correlation functions and  
the transport coefficients. 
In infinite dimensions the vertex corrections to $\Lambda(\omega)$ 
vanish\cite{khurana.90}, so that $\gamma({\bf k},\omega)=1$, and 
once the self-consistency condition is satisfied, the  transport properties 
are determined by the  self-energy of the impurity model.   
From the causality of the problem it follows that  
${\mathrm Im}\ \Sigma_f(\omega^+,T)$ is negative on the real $\omega$-axis. 
At $T=0$, the imaginary part of the self energy of the auxiliary impurity model 
has a maximum at $\omega=0$, such that\cite{hewson} 
${\mathrm Im}\ \Sigma_f(0)=0$,
 ${\mathrm Im}\left[\partial \Sigma_f/\partial \omega\right]_{\omega=0^+}=0$,
and ${\mathrm Im}\left[\partial^2\Sigma_f/\partial \omega^2\right]_{\omega=0^+}<0$.  

The charge-current correlation function  [for the FL at low
temperature where we use Eq.~\eqref{Lambda} instead of 
Eq.~\eqref{eq: corr_fcn_dmft} since it is the dominant 
contribution when $T\rightarrow 0$] follows from the identity 
$G_c({\bf k},\omega^+)G_c({\bf k},\omega^-) =
-A_c({\bf k},\omega)/{\mathrm Im} \ \Sigma_c(\omega^+,T)$, where 
\begin{equation} 
                            \label{A_c_1}
A_c({\bf k},\omega)= -\frac {1}{\pi} {\mathrm Im} \  G_c({\bf k},\omega^+) 
\end {equation}
is the spectral function of conduction electrons. This gives 
\begin{equation} 
                            \label{sigma2}
{\Lambda(\omega,T)}
=
\frac{-e^2}{V\ { {\mathrm Im}\Sigma_c(\omega^+,T)}} 
\sum_{\bf k} {v_{\bf k}}^2 
{ A_c({\bf k},\omega)}  , 
\end {equation} 
and can be further simplified by taking into account that the factor $(-df/d\omega)$ 
restricts the transport integrals to the Fermi window,  $|\omega|\leq k_BT$.  
The  main contribution to transport integrals comes from the {\bf k}-points in 
the vicinity of the renormalized Fermi surface  (FS) and in this narrow region of 
$({\bf k},\omega)$--space we approximate $v_{\bf k}^2$ by its FS average, 
$\langle v_{k_F}^2\rangle$.  This gives the usual result, 
\begin{equation} 
                            \label{sigma_omega}
{\Lambda(\omega,T)}
=
{e^2} \langle{v_{k_F}}^2\rangle {\cal N}_c(\omega) \ \tau(\omega,T) , 
\end {equation} 
where the renormalized {\it c}-DOS is defined by 
\begin{equation} 
                            \label{N_c}
{\cal N}_c(\omega) =\sum_{\bf k}  A_c(\epsilon_{\bf k},\omega),
\end {equation}
 and the transport relaxation time is 
 \begin{equation} 
                            \label{tau}
 \tau(\omega,T)
 = 
 \frac{1}{-  {\mathrm Im} \ \Sigma_c(\omega^+,T) }  .
\end {equation}  
Note,  the approximate FL charge-current correlation function,
given by Eq.~\eqref{sigma_omega},
coincides with the exact DMFT result\cite{freericks.01,freericks.03a} 
only in the limit $T,\omega\to 0$ (the DMFT result has another term, which  
can be neglected as $T\rightarrow 0$, but is important at finite $T$ or for non FL
systems). 

It is a challenge to determine the conductivity and Seebeck coefficient at
low temperature for a FL described by a transport relaxation time given
in Eq.~\eqref{tau}, because the rhs diverges when $\omega$ and $T$ both equal
zero.  Since ${\mathrm Im} \Sigma_f(\omega^+,T)
\approx -|c|(\omega^2+\pi^2 T^2/2)$
for a FL, we see that the proper way to take the limit of $T\rightarrow 0$
is to first consider the limit $\omega\rightarrow 0$ at finite $T$ and then
examine what happens as $T\rightarrow 0$.  
Doing so, will allow for a proper calculation of the
Seebeck coefficient, which is finite, even though it is determined as the
ratio of two integrals, each becoming infinite as $T\rightarrow 0$.

With these ideas in mind, we substitute Eq.~\eqref{sigma_c} for
${\mathrm Im} \ \Sigma_c(\omega^+,T) $ to give 
\begin{equation} 
                            \label{tau_delta}
\tau(\omega,T)
\simeq 
\lim_{\delta\to 0}
\frac{ [\omega - \tilde\epsilon_f(\omega,T)+\mu ]^2+[{\mathrm Im} \ \Sigma_f(\omega^+,T)]^2}
{\delta - {\mathrm Im} \ \Sigma_f(\omega^+,T) V^2 } ,
\end {equation} 
where $ \tilde\epsilon_f(\omega,T)=E_f+{\mathrm Re} \ \Sigma_f(\omega^+,T)$ and 
the  limit is  taken in such a way that $\omega\to 0$ before $\delta\to 0$ 
(i. e., we take $\omega\rightarrow 0$ before $T\rightarrow 0$). 
At low temperature, we do not expect $-\tilde\epsilon_f(\omega,T)+\mu$
to vanish, because it vanishes for the single-band model at half filling,
and we have a multiband model far from half filling.  In the following, we
assume that $-\tilde\epsilon_f(\omega,T)+\mu$ is much larger (in absolute
magnitude) than $\omega$ or $T$ in the low-frequency and low-temperature
regime.  For a given value of $V$, $E_f$ and $n$, we calculate 
$\Sigma_f(\omega)$, $ \tilde\epsilon_f(\omega,T)$, 
${\cal N}_f(\omega)$,  ${\cal N}_c(\omega)$, $n_f$  and $n_c$ by the DMFT 
procedure and find the renormalized $\mu$  from the condition $n_f+n_c = n$.  

The low-temperature Seebeck coefficient is obtained from Eqs.~\eqref{alpha}  
and \eqref{sigma_omega} by the Sommerfeld expansion. It is a weighted sum
of the contribution of all the symmetry channels, which, all of them being equivalent,
is equal to the single channel value. To lowest order: 
\begin{equation} 
                            \label{alpha-tau}
\frac{\alpha(T)}{T}
=
-\frac{\pi^2}{3}\frac{k_B^2 }{|e|}
\left\{
 \frac{1}{ {\cal N}_c (\omega)}\frac{\partial {\cal N}_c (\omega)}{\partial\omega}
 +
 \frac{1}{\tau(\omega)} \frac{\partial\tau (\omega)}{\partial\omega}
\right\}_{\omega,T=0}  . 
\end {equation} 
The quantity   
$\lim_{T\rightarrow 0}\tau_0(T)=\ \lim_{T\rightarrow 0} [\lim_{\omega\to 0} \tau(\omega,T)]$ 
diverges but the expressions given by the ratio of two transport integrals, 
like the one in Eq.~\eqref{alpha}, or the logarithmic derivative in Eq.~\eqref{alpha-tau},
remain finite as the temperature goes to zero. 
Using $ {\mathrm Im} \Sigma_f(\omega)\simeq \omega^2$ at $T=0$ 
we obtain from Eq.~\eqref{tau_delta} 
\begin{equation}
                            \label{partial_alpha-tau}
\left.
 \frac{d}{d\omega} [\ln \tau(\omega)]  
\right|_{\omega=0}    
\simeq
2\left[
\frac{
1-\frac{\partial\Sigma_f(\omega)}{\partial\omega} } 
{\mu -\tilde\epsilon_f(\omega)}  
\right]_{\omega=0}   .
\end {equation}
To find ${\alpha(T)}/{T}$  in Eq.~\eqref{alpha-tau} we have to estimate 
the renormalized position of the {\it f}-states and the renormalized DOS at $\omega,T=0$.


The low-energy densities of states ${\cal N}_c(\omega) $ and ${\cal N}_f(\omega) $ 
are determined by the analytic properties of $A_c(\epsilon_{\bf k},\omega)$ and 
$A_f(\epsilon_{\bf_k},\omega)$,  
considered as functions of the variable $\epsilon_{\bf k}$ and for fixed values 
of  $\omega$, $\tilde\epsilon_f$, $V$, and $\mu$. 
Using the fact that  $A_c(\epsilon_{\bf_k},\omega) $  and $A_f(\epsilon_{\bf_k},\omega)$ 
depend on $\bf k$ only through $\epsilon_{\bf k}$ 
[see Eqs.~\eqref{G_c} and \eqref{G_f} and recall that $\Sigma_f$ is
independent of {\bf k} in DMFT] we write  
\begin{equation} 
                            \label{N_f_1}
{\cal N}_f(\omega) = \int  d\epsilon \ {\cal N}_c^0(\epsilon)  A_f(\epsilon,\omega) ,
\end {equation}
and
\begin{equation} 
                            \label{N_c_1}
{\cal N}_c(\omega) =\int d\epsilon \ {\cal N}_c^0(\epsilon)  A_c(\epsilon,\omega) , 
\end {equation}
where 
${\cal N}_c^0(\epsilon)$ is the unrenormalized density of states of
the conduction electrons (recall $D$ is the half bandwidth of the 
unrenormalized band). 
For small $\omega$, the main contribution to ${\cal N}_f(\omega) $ and 
${\cal N}_c(\omega) $  is coming from the domain of integration 
in which the spectral functions diverge (once again, for the FL at low $T$). 
The condition for  the singular region is obtained from 
Eqs.~\eqref{G_f} and  \eqref{G_c} as 
\begin{equation} 
                                     \label{V}
{\omega-\epsilon_{\bf k}+\mu }
=
\frac{V^2} {\omega -\tilde\epsilon_f(\omega)+\mu} .
\end{equation}
A straightforward calculation gives  on the critical surface 
\begin{eqnarray} 
                                     \label{A_c_4}
A_c(\epsilon,\omega)
&\simeq&
\delta\left({\omega-\epsilon+\mu } - \frac{V^2}{\omega -\tilde\epsilon_f+\mu}\right) 
 \end {eqnarray} 
and
\begin{eqnarray} 
                                     \label{A_f_4}
A_f(\epsilon,\omega)
&\simeq&
\frac{\omega-\epsilon +\mu }{\omega -\tilde\epsilon_f+\mu}
\delta\left({\omega-\epsilon+\mu } -\frac{V^2}{\omega -\tilde\epsilon_f+\mu}\right) , 
\nonumber \\
\end {eqnarray} 
where we used 
$ {\mathrm Im} \Sigma_f(\omega)\simeq \omega^2\ll \delta$ and $\delta^2\ll V^2/D^2$.  
Note, the prefactor of the $\delta$-function in Eq.~\eqref{A_f_4} is positive definite. 
Further note how the argument of the delta function is the same for both the
conduction and the $f$-electrons. But because this argument has a complicated
dependence on frequency, there are actually two roots to the equation where we
set the argument equal to zero.  These two roots comprise the two
distinct quasiparticle bands, which are the physical elementary excitation
bands and are derived in detail below. 

The integrals in Eqs.~\eqref{N_f_1} and  \eqref{N_c_1} are now straightforward to calculate 
and give the zero-temperature results 
\begin{equation} 
                                     \label{N-c-2}
{\cal N}_c(\omega)
=
{\cal N}_c^0
\left(
\omega +\mu-\frac{V^2} {\omega -\tilde\epsilon_f(\omega)+\mu}
\right)  ,  
\end {equation}  
and
\begin{equation} 
                                     \label{N-f-2a}
{\cal N}_f(0)
=
\frac{\mu-\langle \epsilon_{\bf k}\rangle_{FS}} {\mu - \tilde\epsilon_f(0)} {\cal N}_c(0) 
= 
\frac{\mu-\epsilon_{\bf {k}_F}} {\mu - \tilde\epsilon_f(0)} 
{\cal N}_c^0( \epsilon_{ {\bf k}_F}) . 
\end {equation}  
To derive these results, we assumed that $\epsilon_{ {\bf k}}$ varies slowly
on the renormalized Fermi surface defined by ${\bf k}_F$ and replaced 
 $\epsilon_{ {\bf k}}$  in Eq.~\eqref{V} by the FS average  $ \epsilon_{ {\bf k}_F}$.
Note, ${\cal N}_f(\omega)$ and ${\cal N}_c(\omega)$ are implicit functions of 
the degeneracy factor $\cal L$. The total {\it c}- and {\it f}-DOS are obtained 
by multiplying the above expressions by $\cal L$. 
The renormalized densities of states in Eqs.~\eqref{N-c-2} and \eqref{N-f-2a} 
and the general properties of the self energy of the auxiliary single impurity 
Anderson model\cite{hewson}  are sufficient to obtain the FL properties 
of the periodic Anderson model.  


In the limit $T,\omega^+\to 0$, we make a Taylor series expansion 
of $ \Sigma_f(\omega^+)$ truncated to linear order and approximate 
$\omega-[\tilde\epsilon_f(\omega)-\mu] \simeq (\omega-\tilde\omega_f) Z_f^{-1}$, 
where $\tilde\omega_f$ is the renormalized  position of the {\it f}-level which is 
given by the solution of the equation 
$\tilde\omega_f=\tilde\epsilon_f(\tilde\omega_f)-\mu$  
and $Z_f^{-1}={[1-\partial \Sigma_f/\partial \omega]_{\tilde\omega=0}}$.
Using $\tilde\omega_f=[\tilde\epsilon_f(0)-\mu]Z_f$ in Eq.~\eqref{V} gives 
the secular equation for the quasiparticle excitations close to the FS 
\begin{equation} 
                               \label{QP-equation}
(\omega-\epsilon_{\bf k}+\mu )(\omega -\tilde\omega_f) =\tilde V^2            ,                                                
\end {equation}  
where ${\bf k} $ defines the propagation vector of the QP 
and $\tilde V=V\sqrt{Z_f}$ is the renormalized hybridization. 
The two solutions of Eq.~\eqref{QP-equation}  give the QP bands with energy 
$\omega=\Omega^\pm_{\bf k}$ and dispersion\cite{hewson} 
\begin{equation} 
                               \label{QP-dispersion}
\Omega^\pm_{\bf k}=    
\frac{1}{2} 
\left[
(\epsilon_{\bf k}-\mu -\tilde\omega_f)  \pm \sqrt{(\epsilon_{\bf k}-\mu-\tilde\omega_f )^2+4\tilde V^2}
\right]
+\tilde\omega_f  , 
\end {equation}   
where $\Omega^+_{\bf k}$ and $\Omega^-_{\bf k}$ describe the two QP branches 
separated by the hybridization gap $\tilde V_f^2/D$. 
The upper (lower) QP band is very flat for ${\bf k}$ close to the center 
(close to the boundary) of the Brillouin zone (BZ). 
The hybridized FS is defined by the equation $\Omega^\pm_{\bf k} =0$. 

The precise values of the renormalized  quantities ${\bf k}_F$, $ \tilde\epsilon_f(0)$ 
and $\mu$ can only be obtained from the numerical DMFT solution 
but the approximate values can be inferred from the Luttinger theorem 
which requires that the Fermi volume of hybridized electrons is unchanged by 
the Coulomb interaction $U$. For Ce and Eu compounds the condition $U > D$ 
restricts the number of  {\it f}-electrons  at each site to $n_f<1$. 
We assume $n
\leq 2$ and describe the occupied states of the hybridized 
system by the lower QP branch, such that $\mu<0$. 
Choosing, for simplicity, $\tilde\omega_f +\mu \simeq 0$ we obtain from 
Eq.~\eqref{QP-dispersion} that  $|\Omega^\pm_{\bf k}+\mu|>\tilde V^2/D$ 
everywhere in the BZ and  that 
 $\Omega^+_{\bf k} - \Omega^-_{\bf k} \simeq D[1+  2(\tilde V/D)^2] \simeq D$ 
close to the boundary and the center of the BZ. 
For Ce and Eu systems with $n
 <2$ the function $\Omega^-_{\bf k}$ 
changes sign at ${\bf k}_F$ which is not too far from the unrenormalized values, 
i.e.,  ${\bf k}_F$ is close to the zone boundary. 
For this wave-vector we approximate 
$\Omega^+_{{\bf k}_F} - \Omega^-_{{\bf k}_F} 
\simeq\epsilon_{{\bf k}_F}  - \mu \simeq D$ 
and estimate $( \epsilon_{{\bf k}_F}   -\mu) {\cal N}_c(0) \simeq n_c/{\cal L} $.   
For Yb compounds we restrict  $n_f^{h}<1$ and assume that the system 
is more than half-filled,  such that $\mu  >0$. The lower QP branch is 
full and the upper branch is fractionally occupied. 
Choosing again  $\tilde\omega_f +\mu \simeq 0$ we obtain from 
Eq.~\eqref{QP-dispersion} that  ${\bf k}_F$  is close to  the center of the BZ,  
such that $\epsilon_{{\bf k}_F}   - \mu \simeq -D$ and 
$( \epsilon_{{\bf k}_F} -\mu) {\cal N}_c(0) \simeq - n_c/{\cal L} $. 
This shows that 
\begin{equation} 
                                     \label{N-f-3}
{\cal N}_f(0)
= \pm
\frac{n_c}{\cal L} 
\frac{Z_f}{\tilde\omega_f} , 
\end {equation}  
where the upper sign applies to Ce and Eu compounds in which 
the renormalized {\it f}-level is above $\mu$ (${\tilde\omega_f} >0$)
and the lower sign applies to Yb compounds in which 
the renormalized {\it f}-level is below $\mu$ (${\tilde\omega_f}  < 0$).
The relationship between the low-energy scale 
$\tilde\omega_f$ and the specific heat coefficient of the periodic Anderson model 
$\gamma \simeq (\pi^2k_B^2/3) { {\cal L} {\cal N}_f(0) Z_f^{-1} }$ is provided by 
Eq.~\eqref{N-f-3} as $\tilde\omega_f =  (\pi^2k_B^2/3) (n_c/\gamma)$. 

Eqs.~\eqref{N-f-2a} and \eqref{N-f-3} can now be used to eliminate 
$[{\mu -\tilde\epsilon_f(\omega)}]$ from Eq.~\eqref{partial_alpha-tau} 
and write the logarithmic derivative of $\tau(\omega)$ as, 
\begin{equation}
\left.
 \frac{d}{d\omega} [\ln \tau(\omega)]  
\right|_{\omega,T=0}  
\simeq \
\mp \
\frac{  {2 \cal L  N}_f(0)Z_f^{-1}}{n_c } , 
\end {equation} 
where the upper (lower) sign is appropriate for Ce and Eu (Yb) systems. 
Using for $ {\cal N}_c(\omega)$ the result in Eq.~\eqref{N-c-2}, we find 
for a slowly varying $ {\cal N}_0(\omega)$,  
\begin{equation}
\left.
\frac {d}{d\omega}  [\ln  {\cal N}_c(\omega)] \ll \frac{d}{d\omega} [\ln \tau(\omega)]  
\right|_{\omega=0}  , 
\end {equation} 
such that the first term in Eq.~\eqref{alpha-tau} can be neglected, 
and obtain for the low-temperature Seebeck coefficient the FL law 
\begin{equation} 
                            \label{FL-law}
{\alpha(T)}
= 
\pm
\frac{2\gamma T}{|e| n_c} 
=
 \frac{2\gamma T}{|e| n_c} 
\frac{\tilde\omega_f}{|\tilde\omega_f|}
.
\end {equation} 
The initial  thermopower slope is positive when the Kondo resonance 
is above $\mu$, which corresponds to the intermetallic compounds with Ce and Eu ions. 
It is negative  when the Kondo resonance is below $\mu$, which 
corresponds to the intermetallics with Yb ions. 

The above considerations apply to the systems which fluctuate between 
a non-magnetic and a $\cal L$-fold degenerate magnetic state,  
such as heavy Fermions with large Kondo scale and valence fluctuators. 
They  should also apply to heavy Fermions with the CF splitting, because 
at low enough temperatures the excited CF states are unoccupied and the 
magnetic configuration is characterized by the degeneracy of the lowest CF state. 
Using Eq.~\eqref{FL-law}  and the charge neutrality condition $n=n_f+n_c$ 
we find that the reduction of $n_f$ by pressure or chemical pressure enhances 
slightly  the ratio $q=\lim_{T\to 0} \alpha(T)/\gamma T$.  However, these changes 
are small and we expect the  $q$-ratio to be nearly constant for all strongly correlated 
metals described by the Anderson model away from electron-hole symmetry.  

Note, we calculate the transport properties of strongly correlated systems  
following a completely different route than Miyake and Kohno\cite{miyake.05}, 
even though the QP dispersion is exactly the same. 
In Ref.~\onlinecite{miyake.05}, the QP bands are obtained for an effective 
model of hybridized Fermions with renormalized  parameters 
which take into account the on-site correlation.
However, doubly-occupied {\it f}-states are not explicitly excluded 
from the Hilbert space of the effective Hamiltonian and the number 
of {\it f}-electrons is restricted to $ n_f \leq 1$ only on the average. 
In such a Fermionic model, the charge and heat current density operators 
in the QP representation are given by quadratic forms which commute with 
the effective Hamiltonian. The Jonson-Mahan theorem applies and 
the low-temperature thermopower is obtained from the canonical expression
in Eq.~ \eqref{alpha-tau}. The quasiparticle relaxation is due to scattering off 
external impurities, which give rise to a finite residual resistivity, 
and the logarithmic derivatives of the QP density of  states and the transport 
relaxation time are of the same order of magnitude. 

In our approach we assume an infinitely large Coulomb interaction and use 
the QP states which are defined by the canonical transformation which stepwise 
eliminates the hybridization between the f- and the conduction electrons\cite{huebsch.06}. 
The operator form of the effective QP Hamiltonian obtained in such a way 
is quadratic and the dispersion of the QP states is given by Eq.~\eqref{QP-dispersion} 
but the operator algebra is not Fermionic. The  charge and heat current density 
operators are highly non-trivial in the QP representation and it is not clear that 
they satisfy the Jonson-Mahan theorem nor that the canonical expression in 
Eq.~\eqref{alpha} for the Seebeck coefficient holds. 
To avoid these difficulties we  use the QP representation to estimate the 
Fermi momentum of the interacting system but calculate the heat and 
charge currents for the initial model, where the Jonson-Mahan theorem 
is easily proved. We then use the DMFT self-consistency condition to 
relate the width and the position of the Kondo resonance to the renormalized 
{\it c}- and {\it f}-DOS, which allows us to calculate the transport relaxation time. 

{\subsubsection{Anderson model - 'poor man's approach'}\label{PMA}}
At elevated temperature, the FL law  \eqref{FL-law} breaks down and the 
temperature dependence of $\alpha(T)$ cannot be obtained without  
numerical methods. However, the periodic Anderson model in the
presence of nonzero CF splitting or for large degeneracy cannot be solved exactly
by the DMFT mapping and to estimate $\alpha(T)$ we use a 'poor man's mapping'.  
We  assume that the conduction electrons scatter incoherently 
on the 4{\it f} ions and calculate the transport relaxation time in the 
T-matrix approximation. 
We write ${\Sigma_c(k,\omega^+)} \simeq  n_i T_{kk}(\omega^+)$,
where $ T_{kk}(\omega^+)$ is the single-ion scattering matrix  
evaluated on the real axis, and set the concentration of 4{\it f} ions to $n_i=1$.  
Since transport integrals are restricted to the Fermi window, we 
average ${\Sigma_c(k,\omega^+)} $ over the FS and  write
\begin{equation}
                              \label{tau_SIAM}
\frac{1}{\tau(\omega)} = - {\mathrm Im} \ {\Sigma_c(\omega^+)}. 
\end{equation}

In the case of a single scattering channel (no CF splitting) the vertex corrections 
vanish by symmetry and the conduction electron's self energy in 
Eq.~\eqref{tau_SIAM} 
is given by $ {\Sigma_c(\omega^+)}= V^2 G_{f}(\omega^+)$, where $G_{f}(\omega^+)$ is 
the retarded Green's function of the effective $\cal L$-fold degenerate 
single impurity Anderson model.  
In the case when the degeneracy is lifted by the CF splitting the 
vertex corrections do not vanish but we neglect them anyway and use 
\begin{equation} 
                                         \label{sigma_siam_CF}
 {\Sigma_c(\omega^+)}
=
\sum_\Gamma V_{{\bf k} \Gamma_i} G_{\Gamma_I}(\omega^+) V_{\Gamma_i {\bf k}^\prime}.  
\end {equation} 
 $V_{\Gamma {\bf k} }=\left<\Gamma|V|{\bf k}\right>$ is the matrix element 
for the scattering between the conduction state ${\bf k}$ and the  {\it f}-state belonging 
to the irrep $\Gamma$, and $G_{\Gamma}(\omega^+)$  is  the  corresponding 
Green's function of the  single impurity Anderson model with the CF splitting. 
We calculate $G_{\Gamma}(\omega^+)$  by the NCA 
and obtain $\alpha(T)$  from Eqs.~\eqref{alpha} and \eqref{sigma_omega}. 
(For details regarding the thermopower  obtained by the NCA 
see Refs.~\onlinecite{bickers.87} and \onlinecite{zlatic.05}.)  

The single impurity Anderson model which approximates the periodic model 
used in the 'poor man's mapping' is completely different from the auxiliary 
impurity model in the DMFT approach. 
The conduction electrons of the former participate in the charge and 
heat transport and couple to the  {\it f}-states and the external fields. 
The density of these states is parameterized by some simple function 
(square-root or Lorentzian), which is independent of pressure or temperature. 
Furthermore, the hybridization strength, as parameterized by $\Gamma$ also
is independent of temperature, although it might change with pressure.
On the other hand, the conduction states of the auxiliary  impurity model used 
in the DMFT  mapping do not couple to external fields and have no direct physical
meaning. The auxiliary c-DOS usually does change with both pressure or 
temperature (i.e. $\Delta_0\ne \Gamma$ in general). 
 
We expect the 'poor man's mapping' to provide a reliable solution of  the 
periodic Anderson model at temperatures at which  the mean free path 
of conduction electrons is sufficiently short, such that the coherent scattering 
can be neglected. 
Recent solution of the spin-1/2 model obtained by the DMFT+NRG shows\cite{grenzebach.06}  
that the electrical resistance $\rho(T)$ increases rapidly and is very large at the
temperature $T_K$ at which $\alpha(T)$ has a  maximum. For  $T\geq T_K/2$ 
the DMFT+NRG  shows that $\alpha(T)$ is very similar to the 
exact results\cite{costi.94} obtained for the single impurity spin-1/2 Anderson model. 
This indicates that the 'poor man's mapping', which neglects the coherent scattering,  
is reliable above the FL regime and that it can be used to 
approximate $\alpha(T)$ for $T\geq T_K/2$.  
However, the single impurity model has to be solved by  methods which can 
deal with large Coulomb correlation and the CF splitting. 

The experimental results on the heavy Fermion and valence fluctuators 
provide additional support for the single impurity approach.  
The data show that  the residual resistance 
of ternary and quaternary compounds like EuCu$_2$(Ge$_{1-x}$Si$_x$)$_2$, 
CePt$_{1-x}$Ni$_x$ and YbIn$_{1-x}$Ag${_x}$Cu$_4$ grows rapidly with $x$ 
and for $0.3\leq x\leq 0.8$ the mean free path is reduced by disorder  to about 
a single lattice spacing. In these random alloys the electron propagation is 
incoherent even at $T=0$ and they are very well described by the single impurity model. 
At low doping and in stoichiometric heavy Fermion compounds with small $\rho_0$ 
the impurity description breaks down in the low-temperature regime. 
However,  in these systems $\rho(T)$ and $\alpha(T)$ grow rapidly towards room 
temperature (RT) and attain large maxima at $T_K^\rho$ and $T_K $, respectively. 
In the case of the {\it f}-ions with degenerate {\it f}-states or a small CF 
splitting,
the data show\cite{kim.06,kobayashi.03,hossain.04}  $T_K^\rho < T_K  < {RT} $ 
and we find that above $T_K^\rho/2$  there is not much difference between 
$\alpha(T)$ of the stoichiometric compounds and doped systems. 
In the presence of a CF splitting,  $\rho(T)$ has two maxima: 
a low-temperature one at  $T_K^\rho$ and a high-temperature one at $T_{\rho}$.   
The thermopower of these systems exhibits two maxima as well\cite{low-T-max}:  
a low-temperature one at $T_K > T_K^\rho$ and a high-temperature one at $T_{S}$. 
The values of $\alpha(T)$ at  $T_K$ and $T_{S}$ are $\alpha_{K}$ and  
$\alpha_{S}$, respectively.  
Since the mean free path is short for $T\geq T_K/2$,  
it is not really surprising that  the thermopower of periodic systems 
and random alloys have the same qualitative features above the FL regime. 
The experimental data also show that the functional form of $\alpha(T)$ 
is strongly affected by pressure or chemical pressure. 
The fact that all the qualitative features of the pressure-induced 
variations of $\alpha(T)$ are completely accounted for by the 'poor man's mapping'
justifies, a posteriori,  the approximation which neglects the coherent 
scattering of conduction electrons on the lattice of  {\it f}-ions. 

In what follows, we discuss  the temperature and pressure dependence 
of $\alpha(T)$ above the FL regime using the NCA  solution of the asymmetric 
single impurity Anderson model with infinite {\it f-f} correlation. 
We assume that pressure or doping increase the coupling constant $g$ and 
reduce $n_f$ but do not substantially change the CF splitting $\Delta_{CF}$. 
In the limit of large asymmetry and infinite correlation, the single impurity 
Anderson model has a universal low-temperature scale, given by the Kondo 
temperature $T_K$, which is uniquely related to $n_f$. Large degeneracy 
and small CF splitting lead to large $T_K$ which can be further increased by 
reducing $n_f$. A small Kondo scale is found for low degeneracy or large CF 
splitting, which reduces the effective degeneracy of the {\it f}-state; 
the lowest $T_K$ is found for $n_f\simeq 1$. 

We consider first  the case of a $\cal L$-fold degenerate {\it f}-state and 
show the typical NCA  results\cite{bickers.87,zlatic.05,zlatic.05b} obtained 
for $\cal L$=8 in Fig.~\ref{fig:tep}, where $\alpha(T)$ is plotted as a function 
of temperature for several values of $E_f$ and for constant $\Gamma$. 
The thermopower is characterized by the maximum  $\alpha_S$ at 
temperature $T_{S} \simeq T_K$. The Kondo scale of the $\cal L$-fold 
degenerate Anderson impurity model\cite{hewson} is 
$T_K=3\gamma/\pi k_B$, where $\gamma= (\pi^2 k_B^2/3)  {\cal L} \rho_f(\mu) Z_f^{-1}$ 
is the impurity contribution to the specific heat coefficient and 
$ \rho_f(\mu) =(1/ \pi\Gamma) \sin^2  (\pi n_f/\cal L) $ is the {\it f}-DOS 
of a given symmetry at the Fermi level.  Above $T_K$ the thermopower 
decreases with temperature but  the details depend strongly on $n_f$.   
For $n_f\simeq 1$,  the thermopower has a large high-temperature slope, 
changes sign at $T_0 > T_K$ and assumes large negative values above $T_0$. 
For  $0.75 \leq n_f  < 1$, the values of $T_S$ and $T_0$ are increased,  
while  the slope of $\alpha(T)$ above $T_S$ is reduced.
In this parameter range $n_f$ is temperature-dependent 
and to characterize the  system we use  $n_f(T_K)$. 
Eventually, for $n_f < 0.7$,  we still find a shallow maximum of $\alpha(T)$ below RT 
but the high-temperature slope is very small and the sign-change does not occur. 
A similar behavior is obtained if the coupling constant $g$ is reduced by increasing $\Gamma$. 
The thermopower calculated for smaller $\cal L$  has a lower maximum $\alpha_S$ 
which occurs at lower temperature $T_S$\cite{low-T-max}.  

\begin{figure}    %
\center{\includegraphics[width=1 \columnwidth,clip] {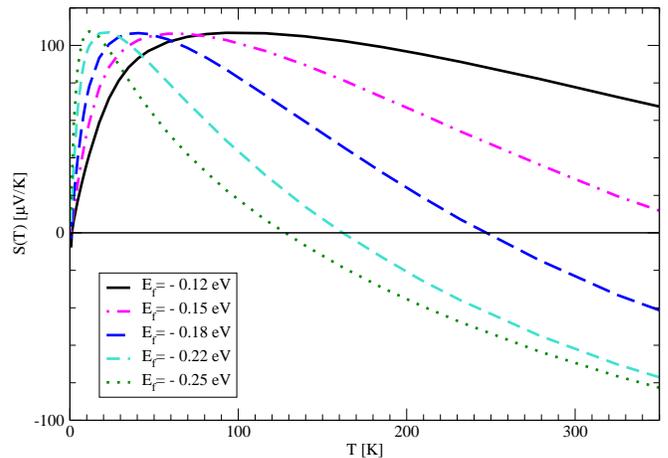}}  
\caption{Thermopower of  the single impurity Anderson model 
of a 8-fold degenerate {\it f}-state is calculated by the NCA for fixed hybridization 
$\Gamma=0.015$ eV and plotted as a function of temperature for several values of $E_f$, 
as indicated in the figure. 
The values of $n_f(T_K)$ are 0.76, 0,81, 0,.86 0.91, and 0.93 for 
$-E_f=$0.12, 0.15, 0.18, 0.22, and 0.25, respectively. 
}
                                                \label{fig:tep_Eu}
\end{figure}

The CF splitting leads to  additional  features which we explain 
by the example of an {\it f}-ion with two CF states separated by $\Delta_{CF}$. 
The respective degeneracies of the ground and the excited state 
are ${\cal L}_1={\cal M}$ and $ {\cal L}_2={\cal M'}$, and we have ${\cal M}+{\cal M}'={\cal L}$. 
The system now has two characteristic low-energy scales: 
 the Kondo temperature $T_K$ and the scale $T_K^{\cal L} \gg T_K$ 
 which comes into play\cite{hanzawa.85} when the excited CF states 
become significantly populated at temperature $T_\Delta\simeq \Delta/2$. 
For  $T\geq T_\Delta$, the thermopower can be  approximated by 
the function $\alpha_{\cal L}(T)$ which describes an effective $\cal L$-fold 
degenerate {\it f}-state with Kondo temperature $T_K^{\cal L}$  and exhibits  
all the typical features discussed in the previous paragraph. 
For $n_f\simeq 1$, the function  $\alpha_{\cal L}(T)$ has a maximum 
$\alpha_S^{\cal L}$ at $T_K^{\cal L}$; above $ T_K^{\cal L}$, $\alpha_{\cal L}(T)$ 
decreases rapidly and changes sign at $T_0^{\cal L} > T_K^{\cal L}$. 
For $0.8 \leq n_f\leq 0.95$, the maximum $\alpha_S^{\cal L}$ is enhanced 
and shifted to higher temperatures, the slope of $\alpha_{\cal L}(T)$ is reduced   
and $T_0^{\cal L}$ is higher.  For $n_f\leq 0.8$, $\alpha_S^{\cal L}$ is further 
enhanced and the sign-change does not occur.  
Of course, in a system with a CF splitting, $\alpha_{\cal L}(T)$ 
has a physical meaning only for $T\geq T_\Delta$. 
At low temperatures, $T  < T_\Delta$, the excited CF states are 
unoccupied and the properties are determined by the lowest CF state 
which is $\cal M$-fold degenerate  (typically, $\cal M \ll \cal L$).  
In the temperature range $T_K/2  < T < T_\Delta$, the behavior of 
a CF split  {\it f} level  is described by an effective ${\cal M}$-fold 
degenerate Anderson model with Kondo scale $T_K$. 
All other parameters being the same,  the main difference between this effective 
model and a simple $\cal M$-fold degenerate model is that  $T_K\gg T_K^{\cal M}$, 
where $T_K^{\cal M}=\lim_{\Delta\to\infty}T_K$. 
The enhancement of $T_K$ with respect to $T_K^{\cal M}$ is 
due to the virtual transitions from the ground to the excited CF states. 
The thermopower  $\alpha_{\cal M}(T)$ of the effective low-temperature model 
exhibits the usual Kondo features. 
For $n_f\simeq 1$,   $\alpha_{\cal M}(T)$ has a maximum at $T_K$ and 
changes sign at $T_0^{\cal M} > T_K$.  Note, for a CF doublet and $n_f\simeq 1$,  
the low-temperature maximum of $\alpha(T)$ could be very small\cite{low-T-max}.  
For $0.7 \leq n_f \leq 0.95$, the maximum of $\alpha(T)$ is enhanced, the slope 
above the maximum is reduced, and the sign-change is shifted to $T_0^{\cal M}\gg T_K$. 
For $n_f \leq 0.7$ the maximum is further enhanced but the sign-change is absent. 
Of course, for $T\geq T_\Delta$ the excited CF states come into play and  
$\alpha_{\cal M}(T)$ ceases to be physically relevant. 
The overall behavior of $\alpha(T)$ for $T\geq T_K/2$ is now easily obtained by 
the interpolation. 

The NCA calculations break down for $T \ll T_K$ but the initial slope of the 
thermopower follows from the Sommerfeld expansion, which gives\cite{costi.94}
\begin{equation}
                              \label{resonant_SIAM}
\lim_{T\to 0}
\frac{\alpha(T)}{T}
= 
\frac{2 \gamma }{|e| n_c}
\cot (\frac{ \pi n_f}{\cal L} )
\end{equation}
for the $\cal L$-fold degenerate Anderson  impurity model. In the case of 
CF splitting, the initial slope  is obtained by substituting  
the effective low-temperature degeneracy of the {\it f}-state into the above
expression, 
i.e.,  replacing  $\cal L$  by the degeneracy of the lowest CF state $\cal M$. 
Since the  on-site correlation is infinitely large and the model is far away from the 
electron-hole symmetry, the initial slope of $\alpha(T)$ is finite, even for the ground state doublet; 
it is positive for Ce and Eu ions, which have an additional electron in the 
magnetic configuration,  and is negative for Yb ions, which are magnetic due to an additional hole. 
However, for ${\cal L}=2$ and $n_f\simeq 1$ the thermopower could be very small.  
To estimate the magnitude of initial slope  we have to take into account  that 
$\gamma$ decreases exponentially as $\cal L$ increases or $n_f$ decreases. 
At constant $\cal L$, we find that $\alpha(T)/T$ decreases with the reduction of $n_f$, 
which is consistent with the NCA results shown in Fig.~\ref{fig:tep_Eu}.
The dependence of $\alpha(T)/T$ on the effective degeneracy of the {\it f}-level 
and $n_f$ is complicated by the fact that the two factors in 
Eq.~\eqref{resonant_SIAM} 
depend strongly on $\cal L$ and $n_f$ but their functional form is completely different. 
Equation \eqref{resonant_SIAM} and the charge neutrality condition 
$n
=n_f+n_c$ show that the reduction of $n_f$ by pressure or chemical pressure 
reduces the ratio $q=\lim_{T\to 0} \alpha(T)/\gamma T$. 
Contrary to periodic systems, in random alloys this reduction can be quite large.  
Note, Eqs.~\eqref{resonant_SIAM} and \eqref{FL-law} give opposite slopes for $q(n_f)$.

The above results, obtained for the CF split Anderson model, explain in simple terms  
the seemingly complicated behavior of  $\alpha(T)$ in many heavy Fermions 
and valence fluctuators with Ce ions. 
The typical NCA results for the model with a ground state doublet at energy $E_f<0$ 
and an excited  quartet at energy $E_f+\Delta_{CF}$ is shown in Fig.~\ref{fig:tep}.  
Various shapes of $\alpha(T)$ are obtained by changing the 
hybridization $\Gamma(p)$, which mimics the effects of pressure or 
chemical pressure. The CF splitting  $\Delta_{CF}$ is the same for all curves 
but pressure increases the coupling constant and the Kondo scale, 
transfers the {\it f}-electrons in the conduction band and reduces, eventually, 
the effective degeneracy of the {\it f}-state. 
As discussed in detail below, the qualitative features at each pressure 
($\Gamma$) are easily explained by approximating $\alpha(T)$ by  
$\alpha_{\cal M}(T)$ at low temperatures and by $\alpha_{\cal L}(T)$ at high temperatures. 
The details of the crossover  at $T\simeq T_\Delta$  require numerical calculations 
but the crossover region is quite narrow and the overall features are easily obtained 
by interpolation. 

\begin{figure}    %
\center{\includegraphics[width=1 \columnwidth,clip] {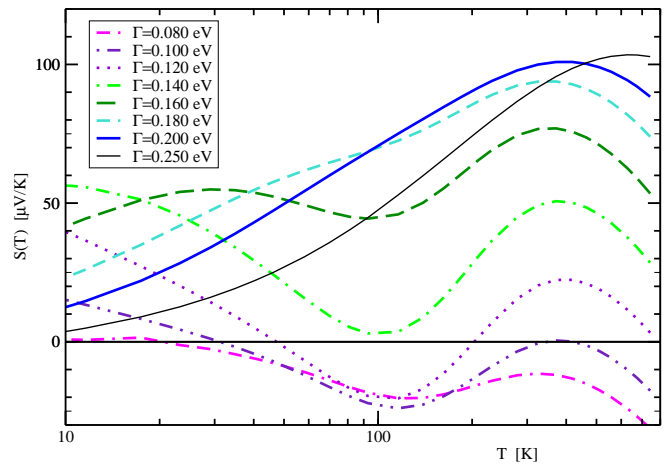}}  
\caption{Thermopower of an {\it f}-ion with the ground state doublet and excited   
quartet calculated by the NCA for  the CF splitting $\Delta=0.07$ eV  is plotted as a 
function of temperature for several values of the hybridization strength $\Gamma$, 
as indicated in the figure. The two bottom curves describe the type (a) Kondo system, 
the third curve from the bottom is type (b), the two middle curves are type (c) 
and the third one from the top describes the type (d) Kondo systems. 
The two upper curves are type (e) and describes valence fluctuators. }
                                                \label{fig:tep}
\end{figure}

Suppose, at ambient pressure $g$ is small enough to give $0.95 \leq n_f\leq 1$.  
The NCA  shows that the spectral function of such a model has well defined  
CF excitations\cite{bickers.87,zlatic.05}  separated by $\Delta_{CF}$ and
 that the low-energy scales satisfy $T_K  \ll  T_K^{\cal L}  \ll   T_\Delta$. 
For $T \leq T_K$  we expect $\alpha(T)$ with a small maximum 
$\alpha_K$ at $T_K$\cite{low-T-max}.  
Above $T_K$,  $\alpha(T)$  decreases and changes sign at $T_0\simeq 2T_K$. 
Nothing particular happens at $T\simeq T_K^{\cal L}$ where the excited CF states  
are still unoccupied. 
At $T_\Delta$,  the excited CF states become thermally populated and the functional 
form of $\alpha(T)$ changes  from $\alpha_{\cal M}(T)$ to $\alpha_{\cal L}(T)$. 
At ambient pressure and $n_f\simeq 1$ we have $\alpha_{\cal M}(T) < \alpha_{\cal L}(T) <0$ 
for $T\simeq T_\Delta$,  i.e., $\alpha_S \leq 0$ at the high-temperature  maximum. 
The thermopower exhibiting these features is shown by the two lowest curves 
in Fig.~\ref{fig:tep} and is classified\cite{jaccard.96,zlatic.05} as type (a). 

If  pressure or chemical pressure increases $\Gamma(p)$ and $g(p)$ such 
that $0.8\leq n_f\leq 0.95$,  the NCA shows that the low-energy  CF excitations 
are still well resolved and $T_K(p)  < T_K^{\cal L}(p)  <  T_\Delta$.  
The  maximum of $\alpha(T)$  at $T_K$ is shifted to higher temperatures 
with respect to the $p=0$ case, we find 
$\alpha_{\cal M}(p,T) > \alpha_{\cal M}(T)$ for $T\geq T_K(p)$,  
and have $ T_0^{\cal M}(p) \gg T_K(p)$.  
The crossover starts at $T_\Delta$ but $T_K^{\cal M}(p)$, $T_0^{\cal M}(p)$ 
and $T_K^{\cal L}(p)$ are now much closer to $T_\Delta$. 
Since $\alpha_{\cal M}(p,T_\Delta)$ and $\alpha_{\cal L}(p,T_\Delta) $ are 
enhanced  with respect to the $p=0$ values, 
pressure brings the two maxima of $\alpha(T)$ closer together,  
shrinks the temperature interval  in which $\alpha(T) <0 $, enhances $\alpha_S$ 
but does not change $T_S$.  These features are demonstrated by the $\Gamma=0.12$ 
curve in  Fig.~\ref{fig:tep}, which is classified as type (b). 

In the pressure range such that $0.75 \leq n_f\leq 0.8$ the 
temperature of the sign-change is pushed up and at large enough 
pressure we get $T_0^{\cal M}(p) \simeq T_\Delta$. 
The crossover starts from $\alpha_{\cal M}(T_\Delta)\geq 0$ and  
$\alpha(T)$ still exhibits two peaks but is always positive. 
These features are demonstrated by the $\Gamma=0.14$  and $\Gamma=0.16$  
curves in  Fig.~\ref{fig:tep}, which are classified as type (c). 
A further increase of pressure  leads to  $0.7\leq n_f\leq 0.75$,  
which shifts $T_K(P)$ close to $T_\Delta$ and the two peaks cannot be resolved 
any more. $\alpha(T)$ exhibits a single peak with a shoulder 
on the low-temperature side, as shown by the $\Gamma=0.18$ curve 
in  Fig.~\ref{fig:tep}, which are classified as type (d). 
Note, as long as the low-energy CF excitations are well defined $\alpha(T)$
has a high-temperature peak  at $T_S\simeq T_\Delta$, which is pressure 
independent, with the maximum value $\alpha_S$, which is pressure-dependent. 

If pressure reduces $n_f$ below 0.7, the NCA calculations show that 
the CF excitations are absent in the spectral function and $\alpha(T)$ 
exhibits the same behavior as a fully degenerate {\it f}-level. 
In this pressure range, $\alpha(T)$  has a single maximum and the difference 
with respect to Kondo systems with well-resolved CF excitations is that
 an increase of pressure shifts $T_S$ to higher temperatures but does not 
 change the magnitude of $\alpha_S$. 
These features are typical of valence fluctuators and are demonstrated 
by the $\Gamma=0.14$  and $\Gamma=0.16$  curves in  Fig.~\ref{fig:tep},   
which are classified as type (e). 
The curves in Fig.~\ref{fig:tep_Eu} are also of this type. They describe a 
system without CF splitting such as Eu intermetallics. The Yb systems are 
characterized by an {\it f}-hole and a qualitative features of $\alpha(T)$ 
are obtained by reflecting ( 'mirror imaging') the curves in  
Fig.~\ref{fig:tep_Eu} and Fig.~\ref{fig:tep} about the temperature axis. 

The NCA solution provides a reliable description of the experimental data 
for $T\geq T_K/2$, where the resistivity is large and the mean free path is short.  
To obtain the overall temperature dependence of the thermopower of the periodic 
systems we interpolate between  the solution obtained by the 'poor man's mapping' 
and the coherent FL solution described by Eq.~\eqref{FL-law}. 
In random alloys the low-temperature behavior is described by a local FL and 
$\alpha(T)$ is obtained by interpolating between Eq.~\eqref{resonant_SIAM} 
and the NCA result.   

\subsubsection{Falicov-Kimball model\label{FK model}}
In heavy Fermions described by the periodic Anderson model, 
the entropy is reduced at the crossover from the high-temperature 
paramagnetic phase to the low-temperature FL phase. In these systems,  
the screening of local moments due to the Kondo effect does not affect $n_f$,  
which is nearly temperature independent. 
However, in some Eu and Yb systems, like EuCu$_2$Ni$_2$ or YbInCu$_4$, 
the magnetic moment disappears due to a temperature-induced change 
in $n_f$, i.e., the entropy is reduced by a valence-change transition.  
These systems can be described by the Falicov-Kimball model\cite{freericks.98} 
which takes a lattice of localized {\it f}-sites, which can be either 
occupied or empty, and conduction states which are delocalized 
via a nearest-neighbor hopping. 
The two types of electrons  interact via a short-range Coulomb interaction 
and share a common chemical potential, which controls the total number 
of electrons $n
=n_c+n_f$. The occupation of the {\it f}-states, which can be 
split into several CF levels, is restricted to $n_f <1$.  
For a given total number of electrons, thermal fluctuations change the 
average {\it f}-occupation by transferring electrons or holes from the conduction 
band to the {\it f}-states and {\it vice versa}\cite{freericks.03a}.
The transport coefficients are obtained in the limit of infinite
dimensions by substituting the exact 
conduction electron Green's function into Eq.~\eqref{eq: corr_fcn_dmft} 
and integrating Eq.~\eqref{alpha} numerically\cite{freericks.01,freericks.03}. 
This procedure allows us to discuss not only the dirty FL regime but also
the metal-insulator transition.

The YbInCu$_4$-like intermetallics are described by the Falicov-Kimball model,
with the interaction large enough to open a gap in the conduction band. We also  
assume that at low temperatures $\mu$ is within the lower (or upper) Hubbard band, 
so that the ground state is metallic. 
Since the model neglects quantum fluctuations, the ground state has no {\it f}-holes 
and the conduction electrons are essentially free, 
such that $ q\simeq 1$.
At finite temperature, the  $\nu$-fold degenerate {\it f}-states become fractionally 
occupied and the additional paramagnetic entropy of these excited states 
competes for the free energy with the excitation energy, the kinetic energy of the 
conduction electrons,  and the interaction energy\cite{freericks.98}. 
This  gives rise to a valence transition at a temperature $T_V$, such that 
a substantial number of electrons (in Eu compounds) or holes (in Yb compounds)  
are transferred from the conduction band to the 4{\it f} ions. 
The onset of the 4{\it f} paramagnetism is accompanied by the reconstruction 
of the interacting density of conduction states and the shift of $\mu$ into the gap. 

We find\cite{freericks.01,freericks.03} that the electrical resistance of the 
paramagnetic phase is large and has a maximum at a temperature $T^*\gg T_V$, 
which is of the order of the gap, or the pseudogap in the density of states. 
The thermopower obtained by the DMFT is weakly temperature dependent and 
its sign depends on the band filling.  The maximum of $\alpha(T)$ is also at  $T^*$.  
The overall entropy of the high-temperature phase is very large due to 
the contribution of local moments and $\tilde q\ll 1$. 
In systems with a valence change transition, $\tilde q$ increases sharply to $q\simeq 1$ 
as temperature is reduced below $T_V$, indicating the onset of the free Fermi gas
phase and the change of the Fermi volume. 
At intermediate temperatures the behavior can be quite 
complex\cite{freericks.01,freericks.03},  because both the degeneracy  
of the {\it f}-states and the number of charge carriers change at $T_V$. 

By choosing the parameters of the model so as to increase the occupancy
of the {\it f}-states one can stabilize the gapped phase, for large Coulomb
repulsion, all the way down to zero temperature\cite{freericks.03}.
Calculating the thermopower for the spinless Falicov-Kimball
model on a Bethe lattice gives\cite{freericks.04}
a thermopower that diverges as $\alpha(T) \simeq \Delta/T$,
where $\Delta$ is the value of the gap.
For an intrinsic semiconductor with a density of states increasing as a power law,
and assuming that the frequency dependent conductivity is proportional to
the density of states, we find\cite{freericks.03}   $\alpha(T)\simeq \ln T$.
In both cases, the  corresponding conductivity decays exponentially,
so that the entropy current density generated by the applied field,
 ${\cal S}(T)=\alpha(T) \sigma(T) (-\nabla \phi)$,  vanishes in the limit $T\to 0$,
as required by the third law of thermodynamics\cite{landau,burns.83}.
This example shows that the value of the $q$-ratio of a correlated insulator
or semiconductor can become very large at low temperatures.

By tuning the parameters of the periodic Anderson model or the Falicov-Kimball model, 
the system can be brought into the vicinity of a quantum critical point, 
where strong non-FL features are expected. 
Such a situation has been considered by Paul and Kotliar\cite{paul.01}, 
who computed the entropy and the thermoelectric power of a system of
quasiparticles scattered by two-dimensional spin-fluctuations.  
They find $\alpha(T) \sim T\ln T$ and ${{\cal S}_N(T)}\sim T \ln T$ but, 
unfortunately, do not discuss the order of magnitude of 
$\alpha(T)/{\cal  S}_N(T)$. 

\section{{Discussion of the experimental data} \label{Discussion}}
In this section, we use the theoretical results obtained for the periodic
Anderson and the 
Falicov-Kimball models to discuss the temperature and doping dependence of 
$\alpha(T)$ and $\alpha/\gamma T$ for several intermetallic compounds with Eu, Ce, 
and Yb ions. In these compounds, chemical substitution modifies the character 
of the ground state, changes the characteristic temperature  and  the low-temperature 
values of $\alpha(T)/T$ and $\gamma$ by an order of magnitude, 
strongly modifies the temperature dependence of $\alpha(T)$,  
but does not significantly change the ratio $\alpha/\gamma T$. 
Our theory explains the universal low-temperature features and shows 
that the observed shapes of $\alpha(T)$ are consistent with the ground 
state properties at each doping level. 
 
\subsubsection{Chemical pressure effects in EuCu$_2$(Ge$_{1-x}$Si$_x$)$_2$
\label{Eu} }
In EuCu$_2$(Ge$_{1-x}$Si$_x$)$_2$ intermetallics, Ge doping increases the lattice
parameter and acts as a negative pressure which reduces the coupling constant 
and makes the system more magnetic\cite{fukuda.03,hossain.04}. 
For $x\simeq 1$, the XPS data indicate a significant mixture of Eu$^{2+}$ and 
Eu$^{3+}$ ions, which is typical  for a valence fluctuator.  
For $x < 1$ the weight of the Eu$^{3+}$ configuration is reduced with respect to that 
of the Eu$^{2+}$ configurations and for $x\simeq 0$ the XPS signal is dominated 
by the magnetic Eu$^{2+}$ ions. 
At the critical concentration $x_c=0.65$, there is a change from a FL to an 
antiferromagnetic (AFM) ground state. Thus, doping changes the magnetic 
character of the Eu ions. 

The doping dependence of the $q$-ratio for $x > x_c$ can be explained by 
the periodic Anderson  model which takes into account the 8-fold  degeneracy 
of the Eu$^{2+}$ ions but neglects the excited magnetic states of the Eu$^{3+}$ configuration. 
In EuCu$_2$Si$_2$, the initial slope of $\alpha(T)$ is  small, 
$\alpha/T=0.64~\mu $V/K$^2$, the specific heat is featureless with 
a small linear coefficient, $\gamma=0.065$ J/K$^2$~mole, 
and the q-ratio\cite{hossain.04} is $q|_{x=1}=0.94$. 
For $0.90 \geq x \geq 0.65$, the $\gamma$ value increases with  Ge-doping 
and for $x=0.7$, we find $\alpha/T=2.86~\mu $V/K$^2$,  
$\gamma=0.226$ J/K$^2$~mole, and $q|_{x=.7}=1.21$. 
The slight enhancement of $q(x)$ obtained for $x_c < x <1$ is most likely 
due to the transfer of electrons from the conduction band into the {\it f}-level 
induced by a negative chemical pressure.  
The observed trend agrees with Eq.~\eqref{FL-law}, which predicts that 
a reduction of the charge carrier density increases the $q$-ratio.  
It seems that $q(x)$  increases more rapidly as we approach the AFM transition 
from the paramagnetic side, but the concentration dependence would have 
to be fine-tuned, before a more quantitative conclusion could be reached.  

The temperature dependence of $\alpha(T)$ at various doping levels is 
consistent with the magnetic character of the Eu ions. 
Since $\rho(T)$ increases rapidly with  doping and temperature, we use 
the 'poor man's mapping' and describe the temperature-dependence of 
the thermopower of EuCu$_2$(Ge$_{1-x}$Si$_x$)$_2$ by the NCA solution 
of the 8-fold degenerate single impurity Anderson model.  
The qualitative features of the solution are shown in Fig.~\ref{fig:tep_Eu},  
where the curves with the highest $T_S$ and lowest $n_f$ corresponds to 
the Si-rich compound. The Ge-doping reduces $n_f$, as indicated in the figure. 
The curves shown in Fig.~\ref{fig:tep_Eu} capture the qualitative features of the 
experimental data but the quantitative analysis has to  take into account the non-resonant 
scattering channels and use the Nordheim-Gorter rule\cite{zlatic.05b}.  This gives  
$\alpha(T)= \alpha_{NCA}(T) \rho_{NCA}(T)/\rho_{tot} +\alpha_0\rho_0/\rho_{tot}$,  
where $\rho_{tot}=\rho_0+\rho_{NCA}$ and $\alpha_0$ is the thermopower due to 
the non-magnetic scattering.  In EuCu$_2$(Ge$_{1-x}$Si$_x$)$_2$, the potential scattering 
is large but $\alpha_0$ is small, and the main effect of the Nordheim-Gorter rule 
is an effective reduction of the magnetic contribution. 

At the Si-rich end, $0.9 \leq x \leq 1$, the experimental data show that  $\alpha(T)$ 
is  always  positive. After an initial linear increase $\alpha(T)$ reaches a broad 
maximum, $\alpha_{S} >  40~\mu $V/K, at a temperature $T_{S}\geq $ 125~K. 
The value of $\alpha_S$ does not change  much,  while $T_{S}$ is reduced 
and the low-temperature slope of $\alpha(T)$ is enhanced with Ge-doping. 
These changes  can be understood in terms of pressure effects  
discussed in Sec. III.2 (see Fig.~\ref{fig:tep_Eu}).
For $x\simeq 1$ the maximum of $\alpha(T)$ occurs at the highest temperatures, 
the decay of $\alpha(T)$ for $T>T_S(x)$ is rather slow, and no sign-change 
is observed, which is consistent with the mixed-valent character of the Eu ions 
and $n_f < 1$.  The negative pressure due to Ge-doping  increases 
$n_f$ and stabilizes the magnetic Eu$^{2+}$ configuration. 
 This shifts the maximum of $\alpha(T)$ to  a lower temperature, 
 makes the slope of $\alpha(T)$ above $T_S(x)$ more negative, 
 and gives rise to the sign-change at $T_0(x)$. 
The data show that Ge doping reduces $T_0(x)$ faster than $T_S(x)$. 
Such behavior also agrees well with the NCA solution of the Anderson model 
and allows us to identify $T_{S}(x)=T_K(x)$ and explain the decrease of $T_0(x)$ 
by the reduction of the coupling constant.

For $x<x_c$, long-range magnetic order occurs at  a Neel temperature 
$T_N(x)$, as indicated by the anomaly in $C_P(T)$ and a discontinuity 
in the slope of $\rho(T)$\cite{fukuda.03,hossain.04}. 
The Neel temperature  increases rapidly as $x$  decreases from $x_c$ to $x=0.5$. 
In this concentration range, the XPS data indicate\cite{fukuda.03,hossain.04} 
a strong mixture  of Eu$^{2+}$ and Eu$^{3+}$ ions at temperatures below $T_N(x)$, 
although the Eu$^{3+}$ component is less pronounced than for $x\geq x_c$.
The maximum of $T_N(x)$ is reached for $x \simeq 0.5 $ and below this concentration,  
$T_N(x)$ decreases slowly as $x$ is reduced\cite{fukuda.03}. 
Note,  below 30\% of Si  the XPS data do not indicate the presence of any 
Eu$^{3+}$ configuration, which indicates  a different magnetic ground state 
around $x\simeq 0$ than around $x\simeq x_c$. 

The magnetic transition is difficult to see in $\alpha(T)$ for $0.6<x \leq x_c$, 
where  Ge-doping increases $T_N$, 
reduces $T_K$, and sharpens the Kondo maximum of $\alpha(T)$.  
However, the maximum of $\alpha(T)$ is still fully developed, $T_K$ is  close to $T_N$, 
and  the only effect of the AFM transition 
on $\alpha(T)$ is a slight  change of slope at $T_N$. 
For $0.5 \leq x \leq 0.6$, negative pressure reduces $T_K(x)$ below $T_N(x) $, 
such that $\alpha(T_N) < \alpha_S$ and  $\alpha(T)$ acquires a cusp, 
as its slope  changes at $T_N$ from negative (for $T>T_N$) to positive (for $T< T_N$). 
In this concentration range, $T_N(x)$ is still rather close to $T_K(x)$ and 
the values of $\alpha(T)$ around $T_N(x)$ are large but the cusp makes 
the overall shape of $\alpha(T)$ quite different  from what one finds in the 
samples with a FL ground state, where $\alpha(T)$ has a rounded maximum 
(see Figs. 4 and 7 in Ref.~\onlinecite{hossain.04}). 
Above $T_N(x)$,  the shape of $\alpha(T)$  looks much the same as in the 
non-magnetic samples well above $T_{S}$ (see Fig. 6 of Ref. \onlinecite{hossain.04}).
Using the single impurity Anderson model  to explain the transport properties 
of the paramagnetic phase of EuCu$_2$(Ge$_{1-x}$Si$_x$)$_2$ allows us to 
infer $T_K(x)$ from the shape of $\alpha(T)$ and the value of $T_0(x)$, 
and compare the concentration dependence of $T_K(x)$ with $T_N(x)$ even 
for antiferromagnetic samples in which $T_K(x)\ll T_N(x)$.
For $x\leq 0.5$,  
the Ge doping reduces $T_K(x)$ much faster than $T_N(x)$,  and the AFM transition 
occurs at a temperatures comparable to $T_0(x)$. The thermopower anomaly at $T_N$ 
is now very weak, because for $T\gg T_K$ the Kondo screening is small and 
$\alpha(T)$ is far away from the Kondo maximum. 

To account for the overall temperature and concentration 
dependence of $\alpha(T)$ in samples with an AFM ground state  and $0.5\leq x\leq x_c$,  
we  note that  the Kondo scale is quite large in this concentration range 
so that the 4{\it f} moments order magnetically before the entropy is 
quenched by the Kondo effect. 
The XPS data indicate  a substantial hybridization 
of the 4{\it f} and conduction states even below $T_N$ and 
the hybridized  {\it f}-electrons contribute to the Fermi volume. 
We assume that the paramagnetic entropy is removed at $T_N$ by an 
anomalous spin density wave (SDW) which gaps a part of the FS. 
Indirect evidence for the SDW transition is provided by the large 
specific heat and small effective moment in the ordered phase. 
Direct evidence by neutron scattering data on EuCu$_2$(Ge$_{1-x}$Si$_x$)$_2$  
is lacking, but the SDW transition has been seen recently in high-pressure 
neutron data\cite{raymond.05} 
on the `reduced-moment'  antiferromagnet CePd$_2$Si$_2$. 
(Unfortunately, the high-pressure thermopower and the specific heat 
data on this compound are not available.)
Note, the maximum of $\gamma(x)$ and the initial slope of $\alpha(T)$ is at $x_c=0.65$, 
while $T_N(x)$ has a maximum at $x=0.5$. The non-monotonic behavior 
of $T_N(x)$ and of the AFM transition below $T_K$ are difficult to explain in terms 
of  the simple Doniach diagram, which  considers the scattering of conduction electrons 
on a lattice of localized spins and predicts that $\gamma(x)$ 
and $T_N(x)$ should peak at $x_c$. 

If we assume that the thermopower  tracks the entropy of the charge carriers,  
the large value of $\alpha(T)$ found for $0.55 \leq x \leq 0.65$ 
points to a large hybridization below $T_N$. 
The small values of $\alpha(T)$ found for $x\leq 0.5$  indicate 
that the coupling between the {\it f} and conduction states is reduced by 
negative chemical pressure. 
At the Ge-rich end, the hybridization and the Kondo coupling seem to be 
small and the {\it f}-electrons are completely localized.  For $x\leq 0.3$ 
the magnetic state involves the unscreened local moments of the Eu$^{2+}$ ions. 
Below $T_N$, the conduction states are effectively free, 
except for scattering off of spin waves. 
The low-temperature entropy is dominated by the linear conduction-electron 
contribution and $\alpha(T)$ is too small to show an anomaly at $T_N$. 

\subsubsection{Chemical pressure effects in CePt$_{1-x}$Ni$_x$ intermetallics 
\label{Ce}}
For our next example, we consider the doping effects on $\alpha(T)$ and 
$\alpha/\gamma T$ in CePt$_{1-x}$Ni$_x$. For $x\geq 0.95$, 
this system  is a valence fluctuator with a FL ground state and for  $x<0.95$ 
a heavy Fermion with a ferromagnetic (FM) ground state\cite{gomez.sal.84,espeso.94,sakurai.02}. 
The temperature and concentration dependence of $\alpha(T)$ in the 
paramagnetic phase show all the typical features of a Ce ion with a CF 
split {\it f}-state, as discussed in detail in Sec. III.2. 
The CF splitting is estimated to be about 200 K and $T_\Delta$ is about  100 K\cite{gomez.sal.84}. 
We account for the observed behavior assuming that the expansion of the volume 
due to Pt doping\cite{gomez.sal.84} reduces hybridization and the coupling constant 
but does not change $\Delta_{CF}$.  
The qualitative features of the thermopower agree with the schematic results 
shown in Fig.~\ref{fig:tep} but for a quantitative agreement  we should use 
the appropriate CF scheme and tune the model parameters. 

For $0.9\leq x < 1$, we find the FL features at  low temperatures, a maximum 
$\alpha_S$ at  $T_S\simeq 120$ K, and a slow decay of $\alpha(T)$ above $T_S$. 
Pt doping shifts $T_S$ from 120 K to 100 K without significant change to $\alpha_S$, 
as described by the type (e) curves in Fig.~\ref{fig:tep}. Such behavior is typical of 
valence fluctuators with a single low-energy scale and $n_f$ too small for the CF 
excitations to appear\cite{alpha_Eu_VF}. 
We estimate the initial slope of $\alpha(T)$ for $x=0.95$ to about\cite{sakurai.02} 
$\alpha/T\simeq 3\ \mu $V/K$^2$ and the specific heat coefficient to 
$\gamma=0.120$ J/K$^2$~mole\cite{espeso.94}, which gives $q|_{0.95}\simeq 2.4$. 

For $0.75\leq x\leq 0.95$ a new feature appears and the thermopower assumes  
first the shape (d) and then type (c). 
In this concentration range, Pt doping reduces $\alpha_S$ but does not change $T_S$, 
which indicates the presence of CF excitations and 
two (or more) low-energy scales.  
The high-temperature one, $T_K^{\cal L}$, characterizes 
a fully degenerate CF state, and the low-temperature one, $T_K$, 
characterizes the CF ground state. 
To explain the data, we assume  $T_K < T_K^{\cal L} < T_0 \simeq T_\Delta $, 
which requires $n_f<1$ (see the discussion in Sec. III.2). 
At  RT all the CF states are equally populated, the  {\it f}-state is effectively 
${\cal L}$-fold  degenerate, and $\alpha(T)$ can be approximated by $ \alpha_{\cal L}(T)$, 
which is a monotonic function with a negative slope. 
As the temperature decreases, the excited CF states depopulate 
and at $T_\Delta$ there is a crossover from the high-temperature regime,  
to the low-temperature one, where the degeneracy of the {\it f}-state 
defines the lowest CF state and $\alpha(T)\simeq \alpha_{\cal M}(T)$. 
Since  $n_f < 1$, $T_0^{\cal M}$ is comparable to $T_\Delta$ and 
the functional form of $\alpha(T)$ does not change much at the crossover. 
Negative chemical pressure reduces $T_K^{\cal L}$ and the (negative) slope 
of $\alpha_{\cal L}(T)$ for $T > T_K^{\cal L}$, such that Pt doping reduces $\alpha_S(x)$. 
(Note, $T_S\simeq T_\Delta \gg  T_K^{\cal L}$.) 
However, $\Delta_{CF}$ is not changed by doping and the position of the 
high-temperature thermopower maximum $\alpha_S$  is always at $T_\Delta$. 
The peculiar feature of CePt$_{1-x}$Ni$_x$ in this concentration range is 
the FM transition which occurs at $T_c\simeq T_K$ and is indicated by the 
discontinuous change in the
slope\cite{gomez.sal.84,sakurai.02}  of $\alpha(T)$ and $\rho(T)$.  
(This feature is quite similar to what one finds  at the AFM transition 
in  EuCu$_2$(Ge$_{1-x}$Si$_x$)$_2$.)
At $x=0.85$ the thermopower and the specific heat data give 
$\alpha/T\simeq 4\ \mu $V/K$^2$\cite{sakurai.02} and 
$\gamma=0.180$ J/K$^2$~mole\cite{espeso.94}, such that $q|_{0.85}\simeq 2.1$. 
Using Eq.~\eqref{FL-law}, one is tempted to associate the reduction of $q(x)$ 
by Pt doping with the transfer of {\it f}-electrons into the conduction band.
However, the onset of the FM transition makes the estimate of the initial thermopower 
and the specific heat slope susceptible to large errors and a quantitative analysis is difficult. 
The low-temperature entropy is now due to the magnetic degrees of freedom 
and is unrelated to the entropy of the charge carriers. 

A further  increase in the Pt concentration continuously reduces $T_K$ and increases $n_f$, 
which rapidly shifts $T_0$ to lower temperature, 
such that $T_0 < T_\Delta$ for $x\simeq 0.5$.  
Once the high- and low-temperature regimes are sufficiently 
far apart, a double-peak structure appears in $\alpha(T)$.
The high-temperature peak remains at $T_S(x)\simeq T_\Delta$  
but $\alpha_S(x)$ is systematically reduced by Pt doping.  
Considered as a function of temperature,  $\alpha(T)$ goes through 
a minimum for $T < T_S$,  and then increases towards $S_K$, as 
predicted by the NCA calculations.  
The minimum of $\alpha(T)$ for $x=0.5$ does not reach negative values, 
as shown by the type (c) curve in Fig.~\ref{fig:tep}. 
For $x < 0.5$ there is a range of temperatures for which $\alpha(T) <0$, 
i.e., the type (b) behavior is obtained.  
The data show that  $T_0(x)$ decreases rapidly with Pt doping 
but the corresponding shift of the low-temperature peak at $T_K(x)$ 
cannot be observed because the full development of this Kondo peak 
is intercepted by the FM transition and a cusp in $\alpha(T)$ 
appears at the transition temperature. (This cusp is similar to 
what is observed in EuCu$_2$(Ge$_{1-x}$Si$_x$)$_2$ for  $x \leq 0.5$.)
In summary, the temperature and doping dependence of the Seebeck coefficient 
of CePt$_{1-x}$Ni$_x$ in the paramagnetic phase are in complete agreement 
with the NCA calculations. 

The onset of the magnetic transition below $T_K$  and the non-monotonic 
concentration dependence of $T_c(x)$ are also difficult to understand in terms of the 
Doniach diagram, obtained by a simple comparison of the Kondo and the RKKY scales. 
The large values of $\alpha(T)$ at the transition and the similarity to the
EuCu$_2$(Ge$_{1-x}$Si$_x$)$_2$ data, can be taken as an evidence 
that the ferromagnetic transition in CePt$_{1-x}$Ni$_x$ is also due to 
an anomalous SDW transition which partly gaps the hybridized FS. 

Similar behavior for $\alpha(T)$ is also seen in many other Ce intermetallics like 
Ce(Pb$_{1-x}$Sn$_x$)$_3$ or Ce(Cu$_{1-x}$Ni$_x$)$_2$Al$_3$\cite{sakurai.05}, 
in which the $\alpha/\gamma T$-ratio is about twice as large as in 
EuCu$_2$(Ge$_{1-x}$Si$_x$)$_2$.  It would be interesting to study 
 $q(x)$ as one approaches the critical concentration from the 
paramagnetic side and determine whether  $q(x)$ exhibits different features above 
the AFM and the FM transitions.

\subsubsection{Chemical pressure effects in YbIn$_{1-x}$Ag${_x}$Cu$_4$ intermetallics 
\label{Yb}  }
Another example of chemical pressure effects is provided by 
YbIn$_{1-x}$Ag${_x}$Cu$_4$ intermetallics\cite{cornelius.97,ocko.01,ocko.04} 
which show anomalous behavior due to the fluctuations of Yb ions between 
the magnetic Yb$^{3+}$ configuration with a single {\it f}-hole 
and non-magnetic Yb$^{2+}$ configurations. 
Indium doping  expands the lattice and increases the weight of Yb$^{2+}$  
with respect to Yb$^{3+}$ configuration\cite{sarrao.96} by transferring 
the  electrons from the conduction band to the 4{\it f}-state.  
This reduces the number of {\it f}-holes and increases the Kondo coupling 
which  makes the compound less magnetic.  However,  the total number of 
conduction electrons is increased and the depletion of the conduction band 
due to chemical pressure is compensated by the substitution of the monovalent 
Ag by the trivalent In. 
 
 We consider, first, the behavior of these compounds in the coherent FL regime. 
 YbAgCu$_4$ is a typical heavy-Fermion with  small characteristic 
 temperature\cite{ocko.04,cornelius.97}, as indicated  by an enhanced 
Pauli-like magnetic susceptibility, large specific heat coefficient and 
a large concentration of Yb$^{3+}$ ions.  
 The XPS data show\cite{dallera.02}  that the number of {\it f}-holes in the ground 
 state is large $n_f^h>0.9$. The thermopower and specific heat data 
 give $\alpha/T=2.2$ $\mu $V/K$^2$ and  $\gamma=0.215$ J/K$^2$~mole  at $T\leq 10$ K, 
 such that $q|_{x=1}=0.98$. 
At the In-rich end ($x=0.4$) the system is a typical valence fluctuator 
with large characteristic temperature, as indicated by the slowly varying 
metallic resistivity,  weakly enhanced Pauli-like susceptibility and 
small specific heat coefficient\cite{zlatic.93b}. 
The XPS data\cite{cornelius.97} indicate $n_f^h < 0.9$, i.e., an increased weight 
of the 4$f^{14}$ configuration. 
The thermopower and specific heat data at  10~K  give  
$\alpha/T=0.36$ $\mu $V/K$^2$ and $\gamma=0.036$ J/K$^2$~mole, 
such that $q|_{x=.4}=0.96$. 
For $0\leq x \leq 0.4$, the values of $\alpha$ and $\gamma$ in the FL regime 
do not show any further changes with doping. 
Describing the Ag-rich compounds by the periodic Anderson model we find 
that the  doping dependence  of $q$ is consistent with Eq.~\eqref{FL-law} 
if we take into account that the Ag--In substitution increases the 
number of conduction electrons despite the charge transfer 
induced by the negative chemical pressure.  
Thus, In doping  transforms the system from a heavy Fermion into 
a valence fluctuator and changes the low-temperature value 
of $\gamma$ and $\alpha/T$ by an order of magnitude but 
has only a small effect on the low-temperature  ratio $\alpha/\gamma T$.
The enhanced thermopower and large specific heat coefficient 
indicate  a large Fermi volume due to the hybridized {\it f}-states. 

The temperature dependence of $\alpha(T)$ is consistent with the ground 
state properties for each value of $x$. 
To start with, notice that the electrical resistance of all these compounds 
increases rapidly with temperature or doping and that above the FL regime 
the mean free path is short enough to justify the `poor man's mapping'. 
(For $0.3\leq x\leq 0.8$, the residual resistivity is large and the single impurity 
approach can be extended down to $T=0$, i.e., the translationally invariant 
FL of the periodic system  can be replaced by the local FL.) 
The properties of the stoichiometric  compound YbAgCu$_4$  
above the FL regime are  well described by 
the NCA solution of a CF split Anderson model with a ground state doublet, 
and excited quartet and doublet states at 4 meV and 7 meV, respectively\cite{dallera.02}.  
Taking the coupling $g= V^2 {\cal N}_0(\mu)/|E_f-\mu|$,  such that $T_K=70$~K, we find 
the ground state value $n_f^{h}\simeq 0.9$ and the temperature variation $n_f^h(T)$ which 
agrees with the  XPS data\cite{dallera.02}. 
For the same parameters the NCA calculations give $\alpha(T)$ with a deep negative 
minimum of about $\alpha_S\simeq - 30 ~\mu$V/K and $T_S\simeq 60$ K. 
The CF splitting is too small (or $g$ is too large) to produce any discernible CF structure 
and the overall shape of  $\alpha(T)$ is characterized by a single deep minimum.  
(Note the `mirror image' analogy with the intermetallic compounds with Ce ions: 
$\alpha(T)$  is described by the `mirror image' of the type (d) curve in Fig.~\ref{fig:tep}.) 
The NCA results agrees with the experimental data on YbAgCu$_4$ which 
show\cite{ocko.04}  $\alpha(T)$ with a broad minimum at about $T_S\simeq 50$ K 
and $\alpha_S\simeq -40 \ \mu$V/K.
Above $T_S$ the thermopower has a a large positive slope, which is 
consistent with $n_f^{h}\simeq 0.9$, and reaches small negative values at RT. 
For $T\gg T_K$ the {\it f}-electrons are localized and contribute a large paramagnetic 
term to the overall entropy (the entropy of the unhybridized conduction electrons is small), 
so that $\tilde q \ll 1$. 

For $0.5 \leq x \leq 1$,  the experimental data show\cite{ocko.04}
that  In doping shifts $\alpha(T)$ to higher temperatures, such that $T_S(x) > T_S(0)$.  
The RT values are reduced  by doping ($\alpha(T)$ is more negative) 
but the bare data show $\alpha_S(x)>\alpha_S(0)$.  
Note, the NCA  calculations refer to the magnetic ion contribution and 
a quantitative comparison would require the Nordheim-Gorter analysis 
which cannot be performed since the absolute values of $\rho_0(x)$ are not available. 
The qualitative features of $\alpha(T)$ can be explained by the Anderson model 
with the same CF level scheme as above but with an enhanced coupling. 
We describe the negative chemical pressure due to In substitution by shifting 
the bare {\it f}-level closer to the chemical potential and keeping 
the hybridization $V$ unchanged. 
(For details of the NCA description of Yb compounds see Ref.~\onlinecite{zlatic.05}.)
Thus, the substitution of Ag by In enhances $g(x)$  and for large enough 
doping we get $g(x)\simeq 1$, which transforms YbIn$_{1-x}$Ag${_x}$Cu$_4$ 
into a valence fluctuator. 
The NCA solution shows that an increase of $g(x)$ enhances $T_K(x)$ and $T_{S}(x)$ 
with respect to the $x=0$ values and reduces the slope of $\alpha(T)$ above $T_S(x)$. 
This makes the  RT values of $\alpha(T)$ more negative, in agreement 
with the experiment\cite{ocko.04}.  Using the `mirror image' analogy with Ce 
compounds,  we see from Fig.~\ref{fig:tep} that In-doping transforms the 
type (d) thermopower of a heavy Fermion with large $T_K$  into the type 
(e) thermopower of a valence fluctuator.

For $x \leq 0.5$, the substitution of monovalent Ag by trivalent In, 
brings $\mu$ and $E_f$ in the vicinity of the band edge $E_c$ which gives rise 
to completely new features.  
In this concentration range  YbIn$_{1-x}$Ag${_x}$Cu$_4$ is a valence fluctuator 
and $n_f^h(T)$ is strongly temperature dependent, such that the temperature-induced 
transfer of {\it f}-holes  in the conduction band can reduce $\mu-E_c$ and $E_f-E_c$  to zero. 
Once $\mu$ is within the gap of the density of states, 
the effective hybridization is switched off and the magnetic moment of the 
{\it f}-ions  cannot be quenched by Kondo screening\cite{fradkin.92,burdin.02}.
Thus, the transition from the low-temperature coherent FL state to the high-temperature 
disordered paramagnetic state cannot follow the usual `Kondo route', 
taken by the heavy Fermions.  
The valence fluctuators like YbIn$_{1-x}$Ag${_x}$Cu$_4$ for $x\leq 0.5$,  
belong to a new class of materials in which the transition between the 
low- and high-entropy phase is driven by the Falicov-Kimball interaction. 
This gives rise, at the temperature $T_V$, to a change in the relative occupancy 
of the {\it f}  and the conduction states and an abrupt modification of the properties of the system.
The valence change transition is clearly seen in the XPS data;  above $T_V$,  
the spectra indicate a stable 4$f^{13}$ configuration of Yb ions and below $T_V$ 
one has a mixture of 4$f^{13}$ and  4$f^{14}$ states\cite{cornelius.97}. 
The magnetic character of the Yb ions changes at $T_V$, as indicated by an 
abrupt change of the susceptibility from Pauli-like to Curie-like\cite{cornelius.97}. 
In the high-temperature phase, the  Curie constant is close to the free ion value 
of Yb$^{3+}$. 
The conduction states are also modified at $T_V$,  as indicated by a drastic change 
of the frequency dependence of the optical conductivity\cite{Hancock.04} and by 
a large increase in the resistivity\cite{sarrao.99}. 
The electrical resistance and the Hall coefficient of the high-temperature phase 
of YbIn$_{1-x}$Ag${_x}$Cu$_4$ are typical of narrow-band semiconductors
or semimetals with a very low carrier density, and neither the transport 
nor the thermodynamic properties show any sign of the Kondo effect. 
The proximity of $\mu$ to $E_c$ is indicated in YbInCu$_4$ 
by the Hall data and band-structure calculations\cite{sarrao.99}. 

The anomalous magnetic response of {\it f}-electrons and the metal-insulator transition 
of the conduction states, which is seen in YbInCu$_4$-like systems,  
are well described by the spin-degenerate Falicov-Kimball model\cite{freericks.98}.  
Performing DMFT calculations for a parameter set which yields   
the valence change transition at $T_V= 50$~K and opens a pseudogap of the order of 
$T^*\simeq 500$ K, we obtain the main features of the magnetic susceptibility,  
the XPS data, and the optical conductivity of YbIn$_{1-x}$Ag${_x}$Cu$_4$ 
at temperatures above $T_V$. 
The calculated thermopower\cite{freericks.01} is of the order of a few $\mu V/K$,
and its sign is either positive or negative, depending on the band filling and 
the shape of the conduction band. 
The proximity of $\mu$ to the pseudogap would lead (in a non-interacting system)
to a shallow minimum of $\alpha(T)$ at a temperature of the order of $T^*$, 
but in an interacting system, the valence-change transition destabilizes 
the semiconducting phase, and gives rise to a discontinuity of $\alpha(T)$ at $T_V$. 
The low-temperature FL state has a large characteristic temperature 
and $|\alpha(T)|$ is a linearly increasing  function of temperature. 
Thus,  a cusp or even a discontinuity appears in $\alpha(T)$ at $T_V$. 
Below $T_V$ the entropy of the system is given by the entropy of the conduction states 
and $q=1$. Above $T_V$ the entropy is dominated by the contribution of the localized, 
paramagnetic states, ${\cal S}\simeq R \ln 8$,  and $\tilde q\ll 1$. The large reduction of 
the $\alpha/{\cal S}$ ratio at $T_V$ is an indication of the reduction of Fermi volume.  

\section{Summary and conclusions}
In this contribution, we have discussed  recent experiments which report an 
universal ratio of the low-temperature thermopower and specific heat of heavy Fermion 
and valence fluctuating intermetallic compounds with Ce, Eu, and Yb ions
\cite{sakurai.93,behnia.04,sakurai.05,sakurai.02,gomez.sal.84,espeso.94,fukuda.03,hossain.04,ocko.01,ocko.04,benz.99,loehenysen.94}.  
The data analysis has shown\cite{sakurai.93,behnia.04} that systems with 
very different values of $\alpha/T$  and $\gamma$ often have similar values 
of the low-temperature ratio $q= N_{A}e {\alpha(T)}/{\gamma T}$. 
Here, we considered in some detail the chemical pressure data on 
EuCu$_2$(Ge$_{1-x}$Si$_x$)$_2$, CePt$_{1-x}$Ni$_x$,  and 
YbIn$_{1-x}$Ag${_x}$Cu$_4$ intermetallics, in which the character of the 
ground state is concentration dependent and $\alpha/T$ and $\gamma$ 
change  by an order of magnitude, while $q(x)$ shows only small, 
but systematic, deviations from universality. 
 
We have first addressed this problem on a macroscopic level and derived the $\alpha/{\gamma T}$ 
ratio from transport equations. Using general thermodynamic arguments, and assuming 
that charge and heat are transported at low temperatures by quasiparticle currents, 
we found $q=N_A /N$, where $N/N_A$ is the effective concentration of the charge carriers. 
However, this derivation also showed that for a general many-body system,
the $q=N_A /N$ law is only approximately valid. We have discussed various possible 
sources for the non-universal behavior and pointed out the difficulties which have  
to be taken into account when comparing the experimental and theoretical results 
for systems which are close to a phase boundary. 

We have then analyzed the low-temperature behavior of the periodic Anderson and 
the Falicov-Kimball models using the DMFT approach and found the universal 
ratio $\alpha/{\gamma T}$. 
We have discussed two possible routes which the system can follow to remove  
the entropy of the paramagnetic states at low temperatures and showed that
the ratio of the thermopower and the entropy, $\alpha/{\cal S}$, tracks the Fermi 
volume of the charge carrying states. 
These results explain the near-universality of the $q$-ratio  observed in 
EuCu$_2$(Ge$_{1-x}$Si$_x$)$_2$, YbIn$_{1-x}$Ag${_x}$Cu$_4$, 
and similar systems with a coherent FL ground state. 
A weak concentration dependence of $\alpha/{\gamma}T$ is most likely 
due to the transfer of the charge from the {\it f}-level to the conduction 
band or vice versa, as described by the expression in Eq.~\eqref{FL-law}. 
The CePt$_{1-x}$Ni$_x$ data also show a slight doping dependence 
but the onset of the ferromagnetic transition precludes a quantitative analysis. 
The calculations for correlated insulators, described by the Falicov-Kimball model,
show that $q(T)$ can assume very large values at low temperature. 

We have also shown, using the 'poor man's mapping' 
which approximates the periodic lattice of 4{\it f} ions by the single impurity Anderson model, 
that the temperature dependence 
of $\alpha(T)$ in EuCu$_2$(Ge$_{1-x}$Si$_x$)$_2$, CePt$_{1-x}$Ni$_x$ and 
YbIn$_{1-x}$Ag${_x}$Cu$_4$\cite{sakurai.93,sakurai.05,sakurai.02,fukuda.03,ocko.01,ocko.04} 
is consistent  with the character of the ground state at each doping level. 
The behavior of $\alpha(T)$ calculated by the NCA for $T\geq T_K/2$ 
captures the main features of pressure and doping experiments on 
heavy Fermions and valence fluctuators. 
Since $\alpha(T)$ does not show much structure for  $T\leq T_K/2$,  
the overall temperature dependence of $\alpha(T)$ is easily obtained, 
interpolating between the FL and the NCA results.
Our calculations explain the drastic doping-induced variation of $\alpha(T)$  
observed in many intermetallic compounds with Ce, Yb and Eu ions. 
They also explain the  pressure-induced crossover from a Kondo system 
with small Kondo temperature due to CF splitting  to a Kondo system 
with large Kondo temperature, and eventually to a valence fluctuator, 
as observed in the resistivity\cite{wilhelm.02} and thermopower 
data\cite{wilhelm.05} on CeCu$_2$Ru$_2$. 
The double-peak structure of $\alpha(T)$ observed in heavy Fermions 
(or the single-peak in valence fluctuators)  is reproduced well by the NCA 
results for the single-ion Anderson model with (or without) CF splitting, 
as shown schematically in Figs.~\ref{fig:tep_Eu} and \ref{fig:tep}. 

In the case of random alloys described by the single impurity Anderson model, 
the zero-temperature limit  of  $\alpha/{\gamma T}$ is given by 
the expression in Eq.~\eqref{resonant_SIAM} which differs by $\cot(\pi n_f/{\cal M})$
from the corresponding expression  in Eq.~\eqref{FL-law} for the Anderson lattice. 
The impurity result  applies to  random alloys obtained by the substitutional 
doping of the rare earth sites,  
like Ce$_{1-x}$La$_x$B$_6$\cite{kim.06,kobayashi.03}, 
Ce$_{1-x}$Y$_x$Cu$_2$Si$_2$\cite{ocko.01a},  
Ce$_{1-x}$La$_x$Cu$_2$Si$_2$\cite{ocko.01b},  
Yb$_{1-x}$Y$_x$InCu$_4$\cite{ocko.03} and similar systems. 
It might also apply to ternary and quaternary  systems with 
one rare earth ion per unit cell but  a short mean free path. 
If the effective degeneracy of the {\it f}-state is changed in random alloys 
by chemical pressure or pressure, the factor  $\cot(\pi n_f/{\cal M})$ should 
lead to observable modifications of the $q$-ratio. However, the experimental 
verification is complicated by the fact that the magnetic and non-magnetic 
contributions to $\alpha(T)$ and $C_V(T)$ are often difficult to separate.  

In summary, our results show that the thermopower is a sensitive probe of the 
low-energy excitations and that the seemingly complicated behavior of $\alpha(T)$ 
can be explained in simple terms.  
It is interesting to note that the expressions in Eqs.~\eqref{resonant_SIAM} and \eqref{FL-law}, 
combined with the charge neutrality condition $n
=n_f+n_c$, give a 
zero-temperature $q$-ratio which is different for periodic systems and random alloys. 
The experimental situation could be clarified  by measuring $\alpha/T$ 
and $C_P/T$ at various hydrostatic pressures on a single stoichiometric 
sample with low residual resistance. 
Pressure can change the effective degeneracy of the ground state, 
increase or reduce the characteristic temperature of a given compound by 
several orders of magnitude and provide a crucial test for the `universality' of the 
$\alpha/{\gamma}T$-ratio. 
Pressure also has a dramatic effect on the overall shape of $\alpha(T)$ and we 
hope that the thermopower and specific heat data of of heavy Fermion 
single crystals taken under hydrostatic pressure will soon be available.
Finally, it would be interesting to follow the behavior of the $q$-ratio as 
a paramagnetic system approaches a magnetic transition or 
a metal-insulator transition gives rise to large changes in the Fermi volume. 


\acknowledgments
This work has been supported by the Ministry of Science of Croatia 
(MZOS, grant No. 0035-0352843-2849), the COST P-16 ECOM project, 
the DFG grant No: Dr274/10-1, and  the National Science Foundation (grant No. DMR-0210717). 
Useful discussions with Z. Hossain, C. Geibel, J. Sakurai, K. Behnia, J. Flouquet,  I. Aviani,  
M. O\v cko and B. Horvati\'c are gratefully acknowledged.


\begin{thebibliography}{8.}
\bibitem{luttinger_note} 
Following Luttinger\cite{luttinger.64}, we use the Coulomb gauge and 
couple the external potential to the charge density. 
The coefficient of the thermal gradient is not calculated directly from the quantum 
mechanical averages (thermal force does not appear in the Hamiltonian) 
but follows in an indirect way\cite{luttinger.64}. 
The perturbation expansion is performed for the Hamiltonian which 
includes a small gravitational field and the coefficient of the thermal gradient 
is  identified with the correlation function multiplying the gradient of the 
gravitational potential.
\bibitem{low-T-max}
In Kondo systems with small $T_K$ the  paramagnetic entropy is removed 
at $T_c>T_K$ by a magnetic phase transition. In such systems the low-temperature 
maximum of $\alpha(T)$  might be concealed and it will emerge only if the FL state 
is restored by pressure or magnetic field but these additional complications are neglected here.
\end{thebibliography}
\end{document}